\documentclass[fleqn,usenatbib]{mnras}
\usepackage{newtxtext,newtxmath}
\usepackage[T1]{fontenc}
\DeclareRobustCommand{\VAN}[3]{#2}
\let\VANthebibliography\thebibliography
\def\thebibliography{\DeclareRobustCommand{\VAN}[3]{##3}\VANthebibliography}

\usepackage{graphicx}
\usepackage{amsmath}

\title[Baryon cycle from SHMR]{Constraining galactic baryon cycle using the galaxy stellar-to-halo mass relations}

\author[Y. Chen et al.]{
Yaoxin Chen$^{1}$, Yingzhong Xu$^{1}$ and  Xi Kang$^{1,2}$\thanks{E-mail: kangxi@zju.edu.cn, chenyaoxin@zju.edu.cn}
\\
$^{1}$Institute for Astronomy, School of Physics, Zhejiang University, Hangzhou 310027, China\\
$^{2}$Purple Mountain Observatory, 10 Yuan Hua Road, Nanjing 210034, China
}

\date{Accepted XXX. Received YYY; in original form ZZZ}

\pubyear{2022}

\begin{document}
\label{firstpage}
\pagerange{\pageref{firstpage}--\pageref{lastpage}}
\maketitle

\begin{abstract}
Galaxies display several well-behaved scaling relations between their properties, such as the star formation rate-stellar mass relation (the main sequence) and the stellar mass-halo mass relation (SHMR). In principle, these scaling relations could imply different star formation histories (SFHs) of galaxies and different constraints on galaxy formation physics. In this paper, we derive the SFHs of galaxies by assuming that they always follow the SHMRs at different redshifts and use an empirical model to constrain key processes in their baryon cycle. It is found that, besides cold accretion due to halo growth, outflow of gas produced by stellar feedback has to be recycled to sustain the derived SFHs of galaxies. The recycled fraction is strongly affected by the baryon fraction in accreted low-mass haloes and the mass loading factor which quantifies the ratio between the galactic outflow rate and star formation rate. Our fiducial model predicts that around $20-60\%$ of outflow is recycled in $\sim0.5-4Gyrs$, while simulations predict a slightly higher recycle fraction and a lower recycle time. We argue that strong constraints on the baryon cycle process can be obtained from future observation of the circum-galactic medium (CGM) of galaxies, such as the gas cooling rate of CGM. We also find that the implied SFHs from the SHMRs indicate that galaxies stay on the main sequences only for part of their lifetimes.
Our model reproduces  the evolution of the mass-metallicity relation as well.
\end{abstract}


\section{Introduction}
\label{sec:intro}
Over the past two decades, huge volume of observation data on properties of galaxies has been obtained from large galaxy surveys, such as the SDSS \citep{2000AJ....120.1579Y}, 2dFGRS \citep{2001MNRAS.328.1039C}, GOODS \citep{2004ApJ...600L..93G}, DEEP \citep{2005ApJS..159...41V,2005ApJ...620..595W}, and COSMOS \citep{2007ApJS..172...38S}. These data reveals various global scaling relations of galaxies and some of them are closely related to star formation process, such as the star formation rate-stellar mass relation (SFR-$M_*$) for star-forming galaxies \citep[e.g.][]{2010MNRAS.405.2279O,2011ApJ...730...61K,2012ApJ...745..149L,2012ApJ...754L..29W,2014ApJ...795..104W,2014ApJS..214...15S,2015A&A...575A..74S}, the mean fraction of cold gas ($f_{gas}\equiv M_{gas}/M_*$) in the interstellar medium (ISM) \citep[e.g.][]{2011MNRAS.415...32S,2017ApJS..233...22S,2014A&A...564A..66B,2015ApJ...800...20G,2017ApJ...837..150S,2018ApJ...853..179T}, the mass-metallicity relation (MZR) \citep[e.g.][]{2004ApJ...613..898T,2008A&A...488..463M,2013ApJ...765..140A,2018ApJ...858...99S,2020MNRAS.491.1427S,2021ApJ...914...19S,2019MNRAS.487.2038C}, and the stellar mass-halo mass relation (SHMR)  \citep[e.g.][]{2000MNRAS.318.1144P,2002ApJ...575..587B,2003MNRAS.340..771V,2003MNRAS.339.1057Y,2006MNRAS.368..715M, 2007ApJ...667..760Z,2010ApJ...710..903M,2012ApJ...752...41Y,2013MNRAS.428.3121M,2020A&A...634A.135G}. These scaling relations are particularly useful to constrain the process of galaxy formation. For example, the SFR-$M_*$ relation and the SHMR can separately predict star formation history (SFH) of a galaxy population \citep[e.g.][]{2012ApJ...745..149L,2009ApJ...696..620C}. The gas fraction in ISM and the mass-metallicity relation are strongly affected by the baryon cycle, such as supernovae feedback, gas accretion, outflow by stellar feedback and metal transport. Modelling those scaling relations and understanding their connections remain to be key questions in current theory of galaxy formation.

In the $\Lambda$ cold dark matter ($\Lambda$CDM) framework, it is generally believed that galaxies form at the centre of halos (and subhalos) and their formation is mainly driven by the cooling and condensation of gas in the centre of dark matter halos \citep{1978MNRAS.183..341W}. Thanks to the advance in numerical methodologies and computing speed, models to follow galaxy evolution in a cosmological context have been developed significantly in the past twenty years, including hydrodynamical simulations \citep[e.g.][]{2015MNRAS.446..521S,2014MNRAS.445..581H,2015MNRAS.454.2691M,2015MNRAS.454...83W,2014MNRAS.444.1518V,2014Natur.509..177V,2018MNRAS.473.4077P} and semi-analytic models (SAMs) \citep[e.g.][]{2006MNRAS.365...11C,2011MNRAS.413..101G,2013MNRAS.434.1531F,2015MNRAS.451.2663H,2020MNRAS.491.5795H,2006MNRAS.370..645B,2012MNRAS.426.2142L,2016MNRAS.462.3854L,2005ApJ...631...21K,2014MNRAS.445..970D,2016MNRAS.461.1760H}.

Undoubtedly, both hydrodynamical simulations and SAMs have achieved great success to match the global properties of galaxies, such as the stellar mass functions (SMF) and some scaling relations mentioned above (for a review, see \citet{2015ARA&A..53...51S}). It is however apparent that these models have some limitations and suffer from uncertainties in the physical ingredients of the modelling. Specifically, hydrodynamical simulations rely on assumptions for sub-grid physics, and are limited by numerical resolutions and expensive computational costs, while the SAMs are also becoming enormously complicated, with considerable number of free parameters describing various physical processes. Therefore, the degeneration of model parameters make it difficult to intuitively understand the influence of different physical processes on the evolution of galaxies in a straightforward way. In addition, most models are often tuned to fit the observational scaling relation of galaxy population at a given epoch, usually the stellar mass function at $z=0$, it is thus not so apparent to infer how galaxy scaling relations are connected with each other and what are their separate constraints on galaxy star formation history.

To investigate if galaxies could evolve along different scaling relations simultaneously and the resulting constraints on their baryon cycle, one firstly has to obtain star formation history of galaxies from their scaling relations. Usually two scaling relations, namely the SHMR and the SFR-$M_*$ relation, can be separately used to derive SFHs of galaxies. Firstly, the SHMRs at different redshifts, combined with the average growth of halo mass, can be used to get SFHs of galaxies. This approach was firstly proposed by \citet{2009ApJ...696..620C} to derive star formation rate of galaxies, which are found to be broadly consistent with the data at $z<1$. Another one is the observed SFR-$M_*$ relation, which can also directly predict the SFHs if galaxies are assumed to follow the SFR-$M_*$ relations at different redshifts \citep{2012ApJ...745..149L}. However, it has been pointed out that the SFHs derived from the SFR-$M_*$ relations are inconsistent with the evolution of galaxy stellar mass function, which on the contrary requires a lower SFR at given stellar mass and a SFR-$M_*$ relation that is steeper than observed \citep{2015ApJ...798..115L, 2016ApJ...817..118T, 2017ApJ...837...27C}. The resulted SFHs from the SFR-$M_*$ relations show that galaxies do not always follow the main sequences at different redshifts.

In this work, we follow \citet{2009ApJ...696..620C} to use the SHMRs to track the evolution of galaxies. The main reason for our choice is that the SHMR has been shown as a very fundamental relation for galaxy evolution. Specifically, once the SHMR is fixed, other galaxy properties, such as the stellar mass functions and spatial clusterings, can be well reproduced \citep[e.g.][]{2010ApJ...710..903M,2010MNRAS.404.1111G,2012MNRAS.422..804K}. We suggest the readers to refer the review paper by \citet{2018ARA&A..56..435W} for progress on relevant studies of the SHMRs. Our work is different from \citet{2009ApJ...696..620C} in two aspects. Firstly, we extend our studies to higher redshift as more data at $z>1$ is available in recent years, and we aim to study if the resulted SFR-$M_*$ relations are consistent with the data across a wide range of cosmic time. Secondly and most importantly, we use the resulted SFHs to constrain galactic baryon cycle using an empirical model of galaxy formation, which is a simplified but clean version of the comprehensive SAM. Similar approach was adopted by some studies, such as \cite{2016MNRAS.455.2592R}. In their work, the SFR for central galaxies can be determined by combining SHMR and the halo mass accretion rate. Then using the empirical model of galaxy formation, with the assumption that the interstellar gas mass is constant for each galaxy, some relations between the net outflow parameter and infall efficiency can be derived. Compared with their work, we pay more attention to the galactic baryon physics, especially the recycle process of the ejected gas.
It is well known that galaxy evolution is mainly governed by some key physical processes, such as fresh gas accretion due to halo growth, star formation from cold gas, galactic wind or outflow by stellar feedback, and recycle of gas outflow  \citep[e.g.][]{2010gfe..book.....M}. Numerical studies have independently achieved strong constraints on these processes, such as gas infall pattern  \citep[e.g.][]{2005MNRAS.363....2K,2009ApJ...694..396B,2009Natur.457..451D,2011MNRAS.417.2982F,2011MNRAS.414.2458V}, galactic outflow rate \citep[e.g.][]{2015MNRAS.454.2691M,2018MNRAS.474.4279M,2020MNRAS.494.3971M} and the wind recycle fraction \citep[e.g.][]{2010MNRAS.406.2325O,2016ApJ...824...57C,2017MNRAS.470.4698A,2019MNRAS.485.2511T,2019MNRAS.490.4786G,2020MNRAS.497.4495M}. We aim to investigate in this work, by requiring our model to follow the SFHs derived from the observed SHMRs at different redshifts, how the parameters on baryon cycle could be constrained and whether they are consistent with results from simulations. 

The structure of this paper is organized as the followings. In Section \ref{sec:method}, we describe the approach to  model galaxy evolution, mainly including two parts: the derivation of galaxy star formation history from the SHMRs at different redshifts and an empirical model to describe the evolution of cold gas mass and chemical enrichment. Our main results of the fiducial model are presented in Section \ref{sec:results_fid}, including the comparison between predicted star formation rate with the observed main sequence (Section \ref{subsec:ms}), the prediction of gas fraction at high redshift (Section \ref{subsec:prediction_of_gas_fraction}), the recycle of gas outflow (Section \ref{subsec:gas_cycle}) and the mass-metallicity relation (MZR) (Section \ref{subsec:MZR}). We shortly discuss the SFHs of massive galaxies and the constraints on stellar feedback in Section \ref{sec:discussion}.
Section \ref{sec:summary} summarizes the main results of our work.

Throughout this paper, we assume a standard flat $\Lambda$CDM cosmology with $\Omega_m=0.315, \Omega_\Lambda=0.685$ and $H_0=67.3 kms^{-1}Mpc^{-1}$, and
a \citet{2003PASP..115..763C} initial mass function (IMF).

\section{Method} \label{sec:method}
The details of our method are described in this section. We start with a brief overview of our procedure in Section \ref{subsec:method_overview}. In Section \ref{subsec:SHMR} and Section \ref{subsec:mass_accretion_his} we present the two main ingredients of the method: the SHMR and the average growth of halo/stellar mass for our sample galaxies. A simple empirical model to describe the key physical process of baryon cycle is introduced in Section \ref{subsec:emp_method}. 

\subsection{Overview} \label{subsec:method_overview}
In this part, we briefly introduce the method to derive the star formation history of galaxies and how to use them to constrain the physical parameters about the baryon cycle. We select $10$ model galaxies with their dark matter halo mass ($M_{h}$) evenly distributed between $10^{11}M_\odot$ and $10^{12}M_\odot$ at $z=0$. Note that in Section \ref{sec:discussion} we will shortly discuss the evolution of massive galaxies, but in most part of this work we focus on galaxies in this mass range for two reasons. The first is that our model neglects merger of dark matter halos (also galaxy mergers) as we try to minimize the uncertainties from modelling galaxy mergers and associated star formation. Previous studies have shown that for galaxies with $M_{\ast} < 10^{11}M_{\odot}$ or $M_h < 10^{12}M_{\odot}$, galaxy mergers contribute very little to the stellar mass growth of galaxies \citep[e.g.][]{2009ApJ...696..620C, 2013MNRAS.428.3121M}. \cite{2009ApJ...696..620C} uses a merger model and a no-merger model to derive the SFR of the galaxies, and find that at least for galaxies with $M_{\ast} < 10^{11}M_{\odot}$ and $z<1$, the results obtained by the no-merger model are more consistent with the observations (see Figure 7 of their work ). The Figure 10 of \cite{2013MNRAS.428.3121M} also shows that for galaxies with $M_h < 10^{12}M_{\odot}$, the fraction of stellar mass growth in galaxies due to mergers is less than $5\%$ at $z>0.5$, although at $z=0$ this fraction increases to $\sim 20\%$.
The second reason is that we focus on the predicted evolution of the SFR-$M_*$ relation, which is basically held for star-forming galaxies with  $M_* < 10^{11}M_{\odot}$. Throughout this work, the $10$ model galaxies are named as $G0$, $G1$, $G2$,...,$G9$ with corresponding halo mass at $z=0$ as  $M_{h0}=1\times10^{11}M_\odot, 2\times10^{11}M_\odot, 3\times10^{11}M_\odot,...,1\times10^{12}M_\odot$, respectively.

The stellar mass of the model galaxies and their evolution are set using the SHMRs. In this work, the growth of halo mass is determined from a Monte-Carlo method, with details to be addressed in Section \ref{subsec:mass_accretion_his}. We  follow \citet{2009ApJ...696..620C} to derive the star formation rate of the galaxies based on their halo mass. For example, for the galaxy $G0$, its progenitor halo mass at a given redshift $z$ is $M_{h}(z)$, we use the SHMR at redshift $z$ (see Equation \ref{eq:SHMR}) to get its stellar mass $M_{\ast}(z)$. By performing the same procedure at a slightly later time-step, say $z+dz$, we can then obtain a new stellar mass $M_{\ast}(z+dz)$ based on its new halo mass, $M_h(z+dz)$ at $z+dz$. We ascribe the stellar mass growth to star formation in the galaxy $G0$, so it is straight to get the evolution of its star formation rate.

Once the star formation history of our model galaxies are obtained, we use a simple empirical model in Section \ref{subsec:emp_method} to constrain the physical parameters of baryon cycle. As we mentioned before, the SHMRs of galaxies at different redshifts have been shown as best benchmarks for galaxy evolution \citep[e.g.][]{2018ARA&A..56..435W},  it has never been clearly shown whether the implied SFHs  can be properly reproduced by a model of galaxy formation. In Section \ref{subsec:emp_method} we then investigate whether the required baryonic processes, such as gas infall rate, recycled gas fraction from the outflow, are reasonable and consistent with simulation constraints. 

\subsection{The stellar mass-halo mass relation: SHMR} \label{subsec:SHMR}
In this paper, we use the latest stellar mass-halo mass relation (SHMR) obtained using the abundance matching technique by \citet{2020A&A...634A.135G}. Although there are another two methods to get the SHMR, namely the halo occupation distribution (HOD) \citep[e.g.][]{2000MNRAS.318.1144P,2002ApJ...575..587B,2007ApJ...667..760Z} and the conditional luminosity function (CLF) \citep[e.g.][]{2003MNRAS.340..771V,2003MNRAS.339.1057Y}, these two approaches are usually applied to data at low redshift as measurements of galaxy clustering are not reliable at high redshift. The abundance matching technique is straightforward to link galaxy with halo by assuming that a galaxy property monotonically relates to a halo property, such as the stellar mass of the galaxy and the dark matter mass of the halo \citep[e.g.][]{2004MNRAS.353..189V}. Introduction on the progress of the SHMR is beyond the scope of this paper, and readers are referred to the review paper by \citet{2018ARA&A..56..435W}.
Recently, \citet{2020A&A...634A.135G} determined the SHMRs up to $z=4$ using the COSMOS data, and they adopted the simple double power-law function proposed by \citet{2010ApJ...710..903M},
\begin{equation}
    \frac{M_*}{M_h}(z)=2A(z)\left[\left(\frac{M_h}{M_A(z)}\right)^{-\beta(z)}+\left(\frac{M_h}{M_A(z)}\right)^{\gamma(z)}\right]^{-1},
	\label{eq:SHMR}
\end{equation}
where $A$ is the normalization at the characteristic halo mass $M_A$, while $\beta$ and $\gamma$ are the slopes of the relation at low- and high- mass ends respectively. In our work, we use the best-fit parameters in Table 4 of \citet{2020A&A...634A.135G}, and we linearly extrapolate their relations to lower halo/galaxy mass if needed. The scatter in stellar mass at given halo mass is a function of halo mass \citep[e.g.][]{2010ApJ...717..379B}, and in our work we follow \citet{2020A&A...634A.135G} to use a typical constant with $\sigma_R = 0.2$ dex to their best fitted SHMRs. 

\subsection{Growth of halo and stellar mass} \label{subsec:mass_accretion_his}

\begin{figure}
    \centering
	\includegraphics[width=\columnwidth]{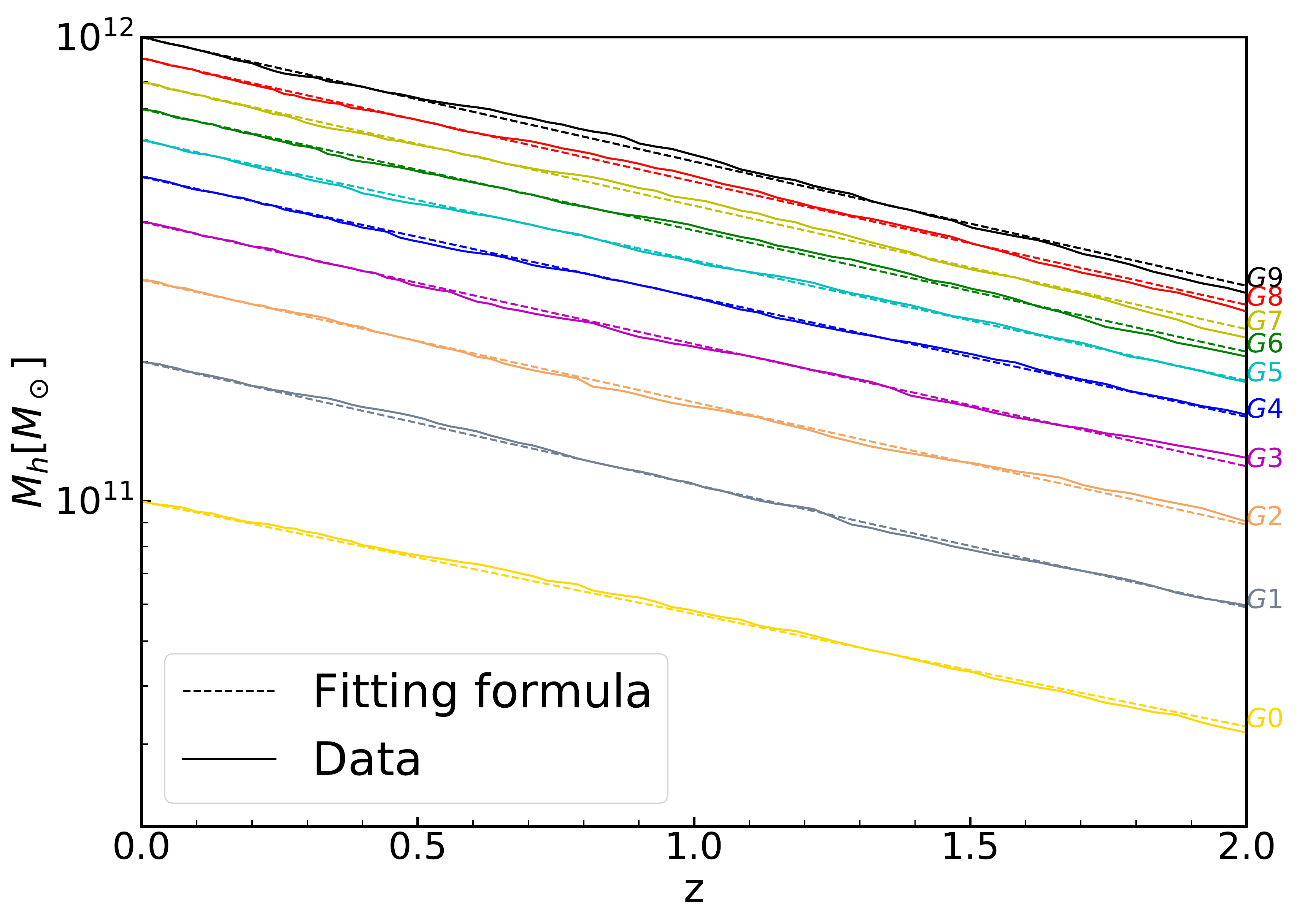}
    \caption{The growth history of halo mass for our ten model galaxies. The solid lines show the mass evolution predicted from the Monte-Carlo merger tree developed by \citet{2008MNRAS.383..557P}. The dashed lines are the best-fittings using Equation \ref{eq:halomassgrowth}.}
    \label{fig:fig1_halo_accretion_history}
\end{figure}

\begin{figure*}
    \centering
	\includegraphics[scale=0.33]{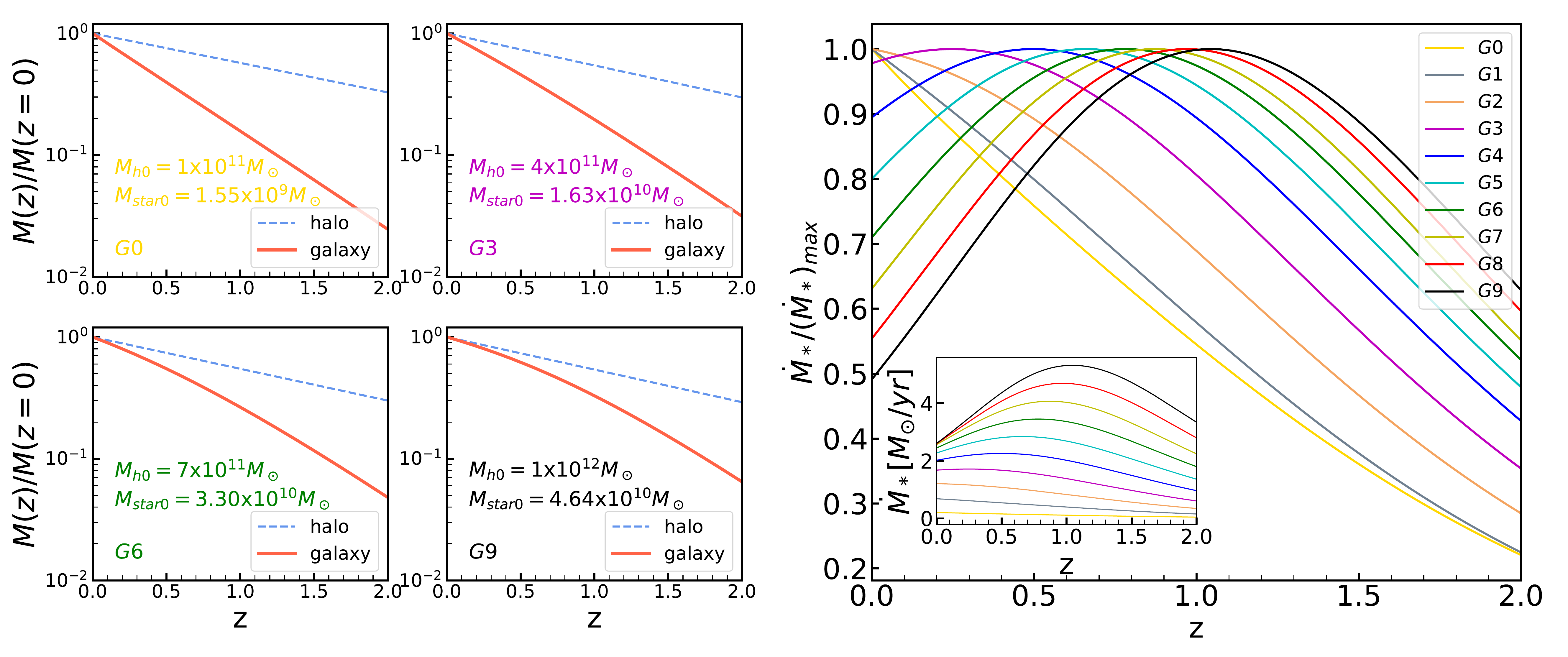}
    \caption{Left panels: the halo and stellar mass evolution for four model galaxies. The mass is normalized using the mass at $z=0$, which is labelled in each small panel. The red lines are for the stellar mass evolution and the dashed lines for the halo mass. It shows that within the mass range of our ten model galaxies halo mass grows faster than the stellar mass, and massive galaxies grow more quickly at high redshift. Right panel: the evolution of star formation rates for the ten model galaxies. Each line is for one galaxy and is normalized by the peak star formation rate of the galaxy during its evolution. The inserted panel shows the absolute star formation rate. This plot shows a strong mass dependence that massive galaxy forms their stars first and has a peak star formation at $z \sim 1$. Low-mass galaxy reaches its star formation peak later or is still not reaching its peak until $z=0$. To make it easier to understand the label, $M_{h0}=(G0,1\times 10^{11}),\ (G1,2\times 10^{11}),\ (G2,3\times 10^{11}),\ (G3,4\times 10^{11}),\ (G4,5\times 10^{11}),\ (G5,6\times 10^{11}),\ (G6,7\times 10^{11}),\ (G7,8\times 10^{11}),\ (G8,9\times 10^{11}),\ (G9,1\times 10^{12})\ M_\odot$.}
    \label{fig:fig2_SFH}
\end{figure*}

The formation history of dark matter halo is obtained using the Monte-Carlo merger tree code developed by \citet{2008MNRAS.383..557P}. For each model galaxy, from $G0$ to $G9$, we produce 100 realizations from the Monte-Carlo code and use the average of these realizations as the halo growth history. We then adopt the simple function of  \citet{2002ApJ...568...52W} to fit the average halo growth history as,
\begin{equation}
    M_h(z)=M_{h0}e^{-\alpha z},
	\label{eq:halomassgrowth}
\end{equation}
where $M_{h0}$ is the halo mass at $z=0$ and $\alpha$ is a free parameter used to fit the data. Although any individual halo trajectory may deviate from this one-parameter function, this one-parameter model can indeed provide a pretty good description of the average halo mass accretion history. Our best fitting result is shown in Figure \ref{fig:fig1_halo_accretion_history} and the best-fitting parameter ranges from $0.56$ to $0.62$, which is consistent with the result of \citet{2002ApJ...568...52W} for halos in this mass range. In fact, the halo mass growth history can also be obtained by other methods, such as integrating the mean mass accretion rate of dark matter obtained from the simulation, that is, integrating $\langle \frac{dM_h}{dt} \rangle$ \citep[e.g.][]{2008ApJ...688..789G,2009MNRAS.398.1858M,2010MNRAS.406.2267F}. After calculation and comparison, we found that there is no obvious difference between our data and those results in our selected halo mass range.

With the evolution of both halo mass (Equation \ref{eq:halomassgrowth}) and the SHMR (Equation \ref{eq:SHMR}) at hand, it is easy to obtain the stellar mass evolution for each model galaxy. As a check for consistency with previous result, in Figure \ref{fig:fig2_SFH} we show the evolution of the normalized halo/stellar mass in the left panels and the star formation rate history in the right panel. The left panels are for four model galaxies with increasing mass from top left to lower right panel. A general trend seen is that within the mass range of our ten model galaxies, the halo mass assembles much earlier than the stellar mass, more noticeable for low-mass galaxy. It can also be seen that galaxy assembles earlier with mass increases. The right panel shows the derived SFHs for the  $10$ model galaxies. It is obvious to find that the peak star formation rate of high-mass galaxy occurs at high redshift, and for low-mass galaxy, such as the galaxy $G0$, its peak star formation has not reached yet at $z=0$. The above results obtained from Figure \ref{fig:fig2_SFH}, whether it is about the evolution of halo/stellar mass or about the star formation rate histories, are in broad agreement with previous related studies \citep[e.g.][]{2009ApJ...696..620C,2013MNRAS.428.3121M,2013ApJ...770...57B}.  
In the following sections, we will then use the derived change rate of stellar mass, $\dot{M}_*(z)$, to constrain the baryonic process.

\subsection{Empirical model for galaxy evolution} \label{subsec:emp_method}
Galaxy formation involves very complicated processes with many details yet to be well understood. Nevertheless, the basic and key processes, such as gas accretion or inflow, cooling of hot gaseous halo, star formation from cold and dense molecular gas, outflow of gas due to stellar wind and supernova feedback, black hole accretion and associated energetic feedback etc, are extensively studied and great progress has been made in past years.  Analytical descriptions of baryonic processes can be easily found from any typical SAM \citep[e.g.][]{2000MNRAS.319..168C,2005ApJ...631...21K,2011MNRAS.413..101G}. 

In this work, we use a simple empirical model to describe the star formation process in our model galaxies.  As mentioned before about our selection of the 10 model galaxies, we neglect galaxy mergers and black hole physics in the model. We mainly include the processes about gas cycle, star formation and its feedback. For star formation rate, we assume a simplified, linear Kennicutt-Schmidt (KS) law \citep{1998ApJ...498..541K}:
\begin{equation}
    SFR(t)=SFE\cdot M_{gas}(t),
	\label{eq:KS}
\end{equation}
where $M_{gas}$ is the cold gas content, and SFE is the star formation efficiency, assumed to be a function of galaxy mass and redshift, and in Section \ref{subsec:prediction_of_gas_fraction} we present how to set its value. 
During the evolution of stars, some mass will return to the inter-stellar medium by stellar wind and supernovae, so the change rate of stellar mass $\dot{M}_*$ in a galaxy is:
\begin{equation}
    \dot{M}_*(t)=(1-R)\cdot SFR(t),
	\label{eq:SF}
\end{equation}
where $R$ is the returned mass fraction from evolving stellar population, which is dependent on the initial stellar mass function (IMF) and can be calculated using the method given by \citet{2016MNRAS.455.4183V}. For our selected Chabrier IMF \citep{2003PASP..115..763C}, we take $R=0.44$.

It has been well known that stellar feedback, via both wind and supernova, plays a very important role in galaxy formation and evolution, and  all modern cosmological hydrodynamical simulations of galaxy formation have included stellar feedback though the details are quite different \citep[e.g.][]{2014MNRAS.444.1518V,2015MNRAS.446..521S,2014MNRAS.445..581H, 2015MNRAS.454...83W}. The major consequence of stellar feedback is the resulted galactic gas outflow that significantly influences the baryon cycle in a galaxy. However, both the relationship between the star formation rate and the outflow mass rate, the so-called mass loading factor, and the fate of the outflow gas are all dependent on the details of modelling stellar feedback in simulations. Even in one particular simulation, the mass loading factor also depends on the properties of the galaxy, and it varies with mass and redshift.

The rate of gas outflow due to stellar feedback is often assumed to be proportional to SFR with a factor depending on galaxy halo mass (or halo velocity) and cosmic time as,
\begin{equation}
    \dot{M}_{gas,out}(t)=\eta(t,M_h)\cdot SFR(t)
	\label{eq:out}
\end{equation}
where $\eta$ is the mass loading factor. Studies using data from cosmological simulations have shown that $\eta$ depends sensitively on the details of feedback in different models \citep[e.g.][]{2015MNRAS.454.2691M, 2016ApJ...824...57C,2017MNRAS.472..949B,2017MNRAS.470.4698A, 2019MNRAS.485.2511T,2019MNRAS.490.3234N, 2020MNRAS.494.3971M}, and we refer the readers to the paper by \citet{2020MNRAS.494.3971M} for comparison of $\eta$ between different cosmological simulations and semi-analytical models. In this work, we use the  fitting formula by \citet{2015MNRAS.454.2691M} to the FIRE simulation \citep{2014MNRAS.445..581H}, and it is dependent on halo mass and redshift as,
\begin{equation}
    \eta=\begin{cases}
    2.9(1+z)^{1.3}\left(\frac{M_h}{M_{h60}}\right)^{-1.10}\quad for\quad M_h<M_{h60},\\
    2.9(1+z)^{1.3}\left(\frac{M_h}{M_{h60}}\right)^{-0.33}\quad for\quad M_h>M_{h60},
    \end{cases}
	\label{eq:massloadingfactor}
\end{equation}
where $M_{h60}$ is the virial mass of a halo with virial velocity ($V_{vir}=\sqrt{GM_h/R_{vir}}$) equals to $60\ km/s$, and $R_{vir}$ is the virial radius radio of the halo within which the mean density is 200 times the critical density of the Universe.

Now we describe the mass change of the gas reservoir due to gas infall and galactic outflow, which is given by the following equation: 
\begin{equation}
    \dot{M}_{gas}(t)=\dot{M}_{gas,in}(t)-\dot{M}_{gas,out}(t)-\dot{M}_{*}(t).
	\label{eq:total}
\end{equation}
Here $\dot{M}_{gas,in}(t)$ is the total gas infall rate, with sources from fresh gas accretion or recycle of outflow. In our model, the gas infall rate is not a free parameter, but is obtained from the above equations. As explained in Section \ref{subsec:mass_accretion_his}, once the evolution of stellar mass change rate, $\dot{M}_*(t)$, is determined for each model galaxy, the evolution of star formation rate $SFR(t)$ can be obtained by Equation \ref{eq:SF}. Then assuming a SFE, the total cold gas content, $M_{gas}(t)$, and outflow rate can be calculated from Equation \ref{eq:KS} and Equation \ref{eq:out} respectively. Thus the gas infall rate can be calculated from Equation \ref{eq:total}.

We will compare the derived total gas infall rate obtained from Equation \ref{eq:total} with model predictions. One main source of gas infall in a galaxy is the fresh gas accretion due to the growth of its dark matter halo. In the classical model \citep[e.g.][]{1978MNRAS.183..341W}, the newly accreted gas is shocked heated to the virial temperature of the halo, and it then cools down via radiative cooling. This kind of process is called as hot mode accretion. Later studies found that the accretion is not always in hot mode. In low-mass galaxies($M_h < 10^{12}M_{\odot}$), newly accreted gas from the external halo is not shock heated, but is directly accreted to the halo center. This cold accretion is rapid and limited by the free-fall time of the halo \citep[e.g.][]{2005MNRAS.363....2K,2006MNRAS.365...11C,2009ApJ...694..396B,2009Natur.457..451D,2011MNRAS.417.2982F,2011MNRAS.414.2458V}. The other source of gas infall is the re-accretion/recycle of the outflow. Recent studies using hydrodynamical simulation of galaxy formation have shown that the fraction of recycled gas from outflow is dependent on the properties of the outflow, and it is diverse among different simulations \citep[e.g.][]{2010MNRAS.406.2325O,2016ApJ...824...57C,2017MNRAS.470.4698A,2019MNRAS.485.2511T,2019MNRAS.490.4786G,2020MNRAS.497.4495M}. In Section \ref{subsec:gas_cycle} we will compare our predictions on gas recycle with simulation results.

We also predict how metals evolve in a galaxy. The production of metal is accompanied by the formation and evolution of stars. Then newly produced metal is returned to the ISM by stellar feedback or stellar wind, and it is assumed to uniformly mix with the ISM. A yield parameter $y_Z$ is often used to characterize this process, which is the ratio of the total mass of metals that a stellar population releases into the ISM and the remained stellar mass of that stellar population. This can be intuitively understood as
\begin{equation}
    y_Z=\frac{\dot{M}_{new\_metal}}{(1-R)SFR}=\frac{\dot{M}_{new\_metal}}{\dot{M}_*},
	\label{eq:yield}
\end{equation}
where the $\dot{M}_{new\_metal}$ is the metal mass growth rate in ISM due to the metal production. Obviously, the value of $y_Z$ is also dependent on the selected IMF. For Chabrier IMF \citep{2003PASP..115..763C}, we take $y_Z$ as 0.06 from the work by \citet{2016MNRAS.455.4183V}. For more definition or calculation details of $R$ and $y_Z$, we refer readers to \cite{2016MNRAS.455.4183V}.
The metallicity of the ISM is defined as,  
\begin{equation}
    Z(t)=\frac{M_Z(t)}{M_{gas}(t)}.
	\label{eq:metallicity}
\end{equation}
Here ${M_Z(t)}$ is the total metal content in the ISM. The metal will also be reduced by outflow or diluted by the accretion of cold gas, so it can be calculated as,
\begin{equation}
    \dot{M}_Z=(-Z(t)+y_Z)\cdot \dot{M}_{*}(t)-Z(t)\cdot \dot{M}_{gas,out}+Z_{in}\cdot \dot{M}_{gas,in}(t).
    \label{eq:metalmass1}
\end{equation}
Here we assume the metallicity of new infall gas is $Z_{in}$ and it is assumed to be primordial with $Z_{in}=10^{-4}$ in our fiducial model, but we will also show predictions of the mass-metallicity relation with different $Z_{in}$ in Section \ref{subsec:MZR}.

\section{Predictions of the fiducial model} \label{sec:results_fid}

In this section we present results from our fiducial model. In Section \ref{subsec:ms}, we show the trajectories of model galaxies on the main sequence relations at different redshifts, and we predict the gas fraction at high redshift in Section \ref{subsec:prediction_of_gas_fraction}. Constraints on the process of gas recycle are shown in Section \ref{subsec:gas_cycle}. We present the evolution of the mass-metallicity relation (MZR) in Section \ref{subsec:MZR}.

\subsection{Trajectories of model galaxies on the Main Sequence (MS) relations} \label{subsec:ms}
\begin{figure*}
    \centering
	\includegraphics[scale=0.177]{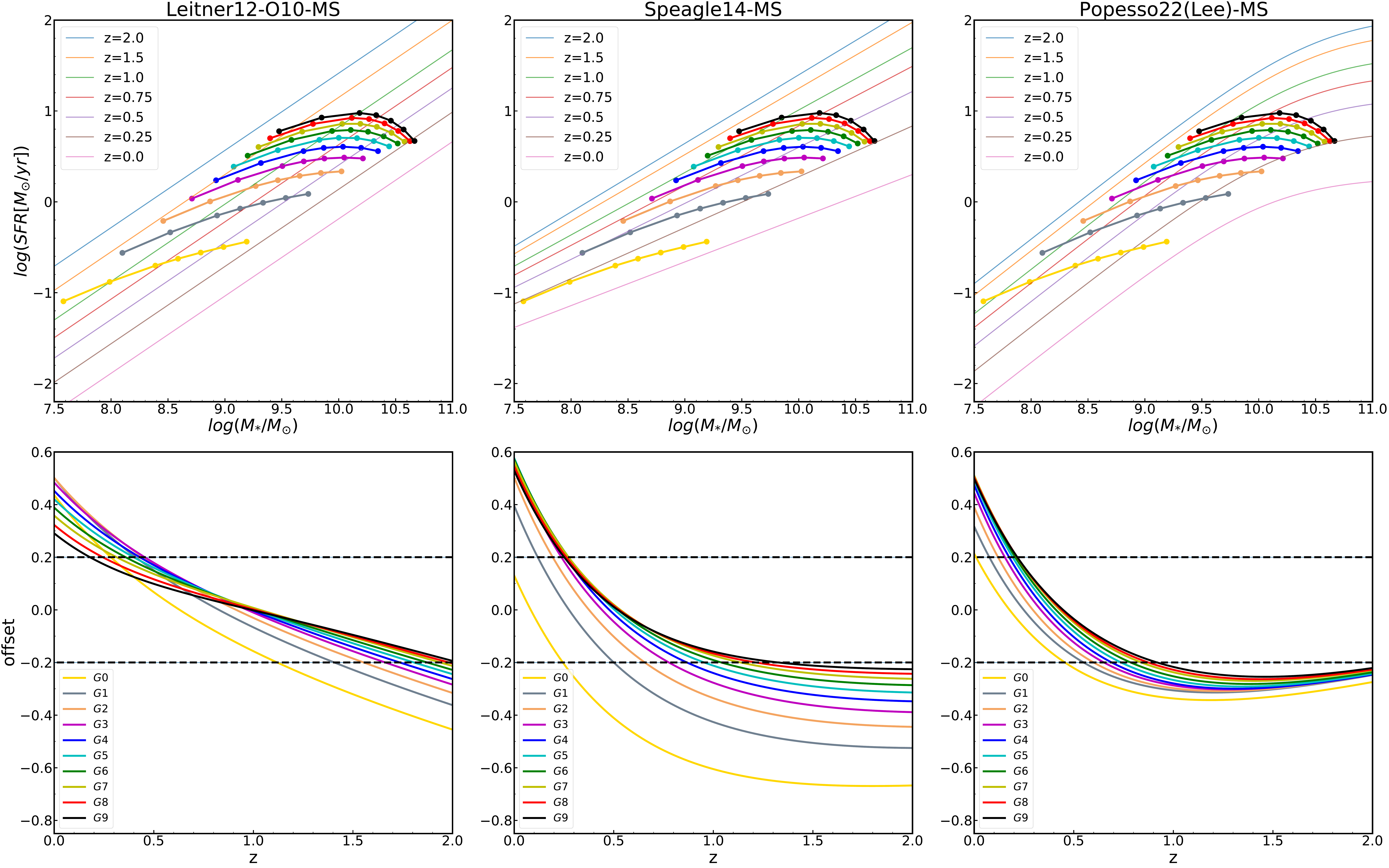}
    \caption{The evolution of model galaxies on the Main Sequence diagram, or the SFR-$M_{\ast}$ relation. Upper panels: the curved lines are the trajectories of the ten model galaxies (each line for one galaxy, see label in lower right panel), and the circles along each curved line show the star formation rate and stellar mass at several redshifts between $z=0$ and $z=2$, indicated by the label in upper panels. The straight color lines are the best-fitting relations to the observational data:  \citet{2012ApJ...745..149L} (left panel), \citet{2014ApJS..214...15S} (middle panel) and \citet{https://doi.org/10.48550/arxiv.2203.10487} (right panel) at different redshifts, also indicated by the label in each panel. Lower panel: the offset between the predicted star formation rate and the best-fitting Main Sequence relations. It is seen that a galaxy is not always a star forming galaxy. For galaxies in the mass range ($10^{11}-10^{12}M_{\odot}$), they have lower star formation rate than the data at high redshift, but are above the Main Sequences at $z \sim 0$. To make it easier to understand the label, $M_{h0}=(G0,1\times 10^{11}),\ (G1,2\times 10^{11}),\ (G2,3\times 10^{11}),\ (G3,4\times 10^{11}),\ (G4,5\times 10^{11}),\ (G5,6\times 10^{11}),\ (G6,7\times 10^{11}),\ (G7,8\times 10^{11}),\ (G8,9\times 10^{11}),\ (G9,1\times 10^{12})\ M_\odot$.}
    \label{fig:fig3_MS}
\end{figure*}

Observational studies have found a tight relation between SFR and stellar mass of star-forming galaxies (SFR-$M_*$), often called the "Main Sequence (MS)", and this relation is observed to hold from $z=0$ to $z=6$ \citep[e.g.][]{2010MNRAS.405.2279O,2011ApJ...730...61K,2012ApJ...745..149L,2012ApJ...754L..29W,2014ApJ...795..104W,2014ApJS..214...15S,2015A&A...575A..74S, https://doi.org/10.48550/arxiv.2203.10487}. As the MS relation is a snapshot of star forming galaxies at one cosmic epoch, one may wonder how a given galaxy evolves along the MS relations at different redshifts. One simple assumption is that  "A star-forming galaxy is always a star-forming galaxy", and it then predicts a unique SFH of galaxy using the MS relations at different redshifts. This technique is called as the "Main Sequence Integration (MSI)" \citep[e.g.][]{2012ApJ...745..149L,2017A&A...608A..41C}. Here we show the trajectories of our model galaxies on the MS diagram and investigate whether they agree with the MSI scenario. 

Before showing the trajectories of model galaxy in the MS diagram, one should bear in mind two facts. The first is that the MS relation is only for star forming galaxy, and quenched galaxies are usually not included. The second is that measurements of both star formation rate and stellar mass are tricky, as they depend on star formation rate indicator and techniques to derive the stellar mass of galaxies.  To ensure a fair comparison with data, here we use the best-fitting MS relations from three studies, which are \citet{2012ApJ...745..149L}, \citet{2014ApJS..214...15S}, \citet{https://doi.org/10.48550/arxiv.2203.10487}.  
In \citet{2012ApJ...745..149L}, the fitting was done separately to the radio survey data from \cite{2011ApJ...730...61K} (hereafter K11) and the IR survey data from \cite{2010MNRAS.405.2279O} (hereafter O10). Here we use their fitting result for O10 data. Although for O10 data, the fitting is only for galaxies with stellar mass of $\sim 10^{11}M_\odot$, compared with the fitting results of K11 data, the fitting result using O10 data is more in line with the latest observational result at low mass end. \citet{2014ApJS..214...15S} converted observational data from 25 studies to a common set of calibrations and gave the best-fitting MS relations within the fitted stellar mass range of $10^{9.7}-10^{11.1}M_\odot$. A recent work by  \citet{https://doi.org/10.48550/arxiv.2203.10487} also did similar analysis by combing data from literatures and obtained the best-fitting MS relations in the widest range of redshift ($0<z<6$) and stellar mass ($10^{8.5}-10^{11.5}M_\odot$), and we plot one set of their best fitting formula in the right panel of Fig.3.
\begin{figure*}
    \centering
	\includegraphics[scale=0.29]{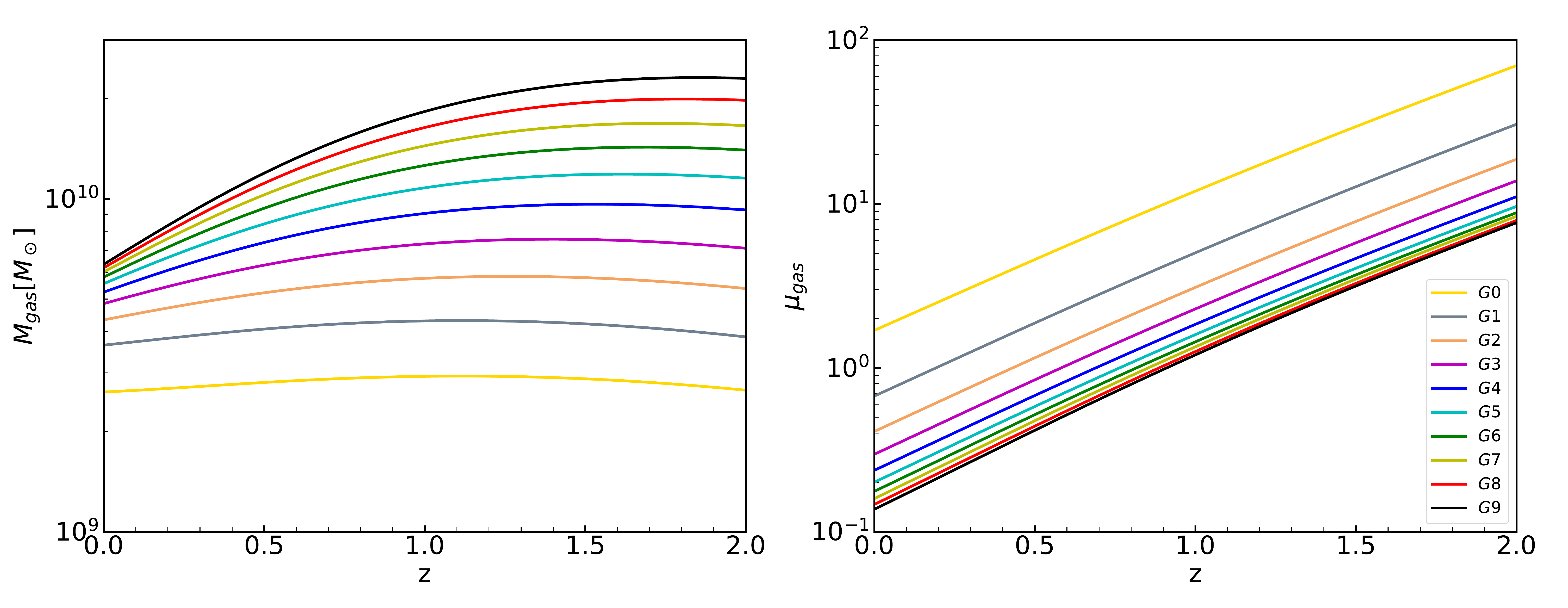}
    \caption{The predicted evolution of cold gas mass (left panel) and the cold gas fraction, defined as  $\mu_{gas}=\frac{M_{gas}}{M_{*}}$ (the right panel) for the ten model galaxies from our empirical model. It can be found that for low-mass galaxies, their gas mass remain almost unchanged since $z=2$ and they were very gas rich in the past, with gas mass about 10 times higher than stellar mass at $z=2$. To make it easier to understand the label, $M_{h0}=(G0,1\times 10^{11}),\ (G1,2\times 10^{11}),\ (G2,3\times 10^{11}),\ (G3,4\times 10^{11}),\ (G4,5\times 10^{11}),\ (G5,6\times 10^{11}),\ (G6,7\times 10^{11}),\ (G7,8\times 10^{11}),\ (G8,9\times 10^{11}),\ (G9,1\times 10^{12})\ M_\odot$.}
    
    \label{fig:fig4_fgasevolution}
\end{figure*}
In the upper row of Figure \ref{fig:fig3_MS}, we show the evolution of the ten model galaxies on the MS diagram, and we compare them with data from the aforementioned three observational work in the left, middle and right panels, respectively. In each panel, the straight color lines are the best-fitting MS relations to the data at different redshifts, and the curved color lines are the trajectories of the ten model galaxies. Each curved color line is for one galaxy (from $G0$ to $G9$, see labels in the lower row) and the circles along the line mark the positions of the galaxy on the MS diagram at different redshifts. For example, the curved yellow line (the lowest line in each panel) is for the galaxy $G0$, and the most left circle show its stellar mass and star formation rate at $z=2$, and the next circle shows the position at $z=1.5$, and so on. It is seen that, compared with data in any of the three panels, our model galaxies are below the MS relations at high redshift and are above the MS relations at low redshift. 

The derivation of the model galaxies from the MS relation can be more clearly seen in the lower row of Figure \ref{fig:fig3_MS} where we show the offset between our predicted star formation rate and the observed MS, as a function of redshift. As the typical scatter of the observed MS relation is around $0.2$ dex with weak dependence on galaxy mass, we plot two horizontal lines to indicate the scatter. It is found that our results agree better with the MS relations from \citet{2012ApJ...745..149L}, and it shows that most of our model galaxies selected at $z=0$ were star forming galaxies between $z=1.5$ and $z=0.5$, and their star formation rates were close to the MS relation. Below $z=0.5$ the model galaxies are still above the MS relations. Larger deviation is seen in the left panel where it shows that our model galaxies all have star formation rate lower than the data at $z>1$. The results show that a star forming galaxy is not always a star forming galaxy, and it remains on the MS for only a limited lifetime.

The predicted lower star formation rate than the data at $z>1.5$ agrees with previous conclusion using similar methods \citep[e.g.][]{2009ApJ...696..620C, 2015ApJ...798..115L}. In fact, similar discrepancies are also seen in SAMs or hydrodynamical simulations, that is, at middle and high redshifts, the main sequence obtained in theoretical models is systematically lower than the observed MS by $\sim 0.2-0.5$ dex \citep[e.g.][]{2010ApJ...713.1301K,2014MNRAS.444.2637M,2015MNRAS.447.3548S,2015MNRAS.450.4486F,2016ApJ...817..118T,2019MNRAS.485.4817D}.
It is recently suggested that the tension could be alleviated if similar methods are used to both data and model galaxy to derive the MS relations \citep{2020MNRAS.492.5592K}. \citet{2021arXiv211004314L} also show that, if a more flexible neural network is adopted to process the data, the normalization of the MS relation can be lowered by $\sim 0.2-0.5$ dex, which is consistent with predictions from theoretical models of galaxy formation.

Before we conclude the discussion on the MS relation, we note that here we only show the trajectories of our model galaxies with halo mass lower than $10^{12}M_{\odot}$. It is well realized that low-mass galaxies form later and are now in the process of rapid star formation. This is why we found they are all above the MS at $z=0$. In Section \ref{sec:discussion} we will show the MS relations of more massive galaxies for comparison, and we will see that massive galaxies are basically quenched with star formation rate lower than the MS relations at $z \sim 0$.

\subsection{Evolution of the cold gas fraction} \label{subsec:prediction_of_gas_fraction}

\begin{figure}
    \centering
	\includegraphics[width=\columnwidth]{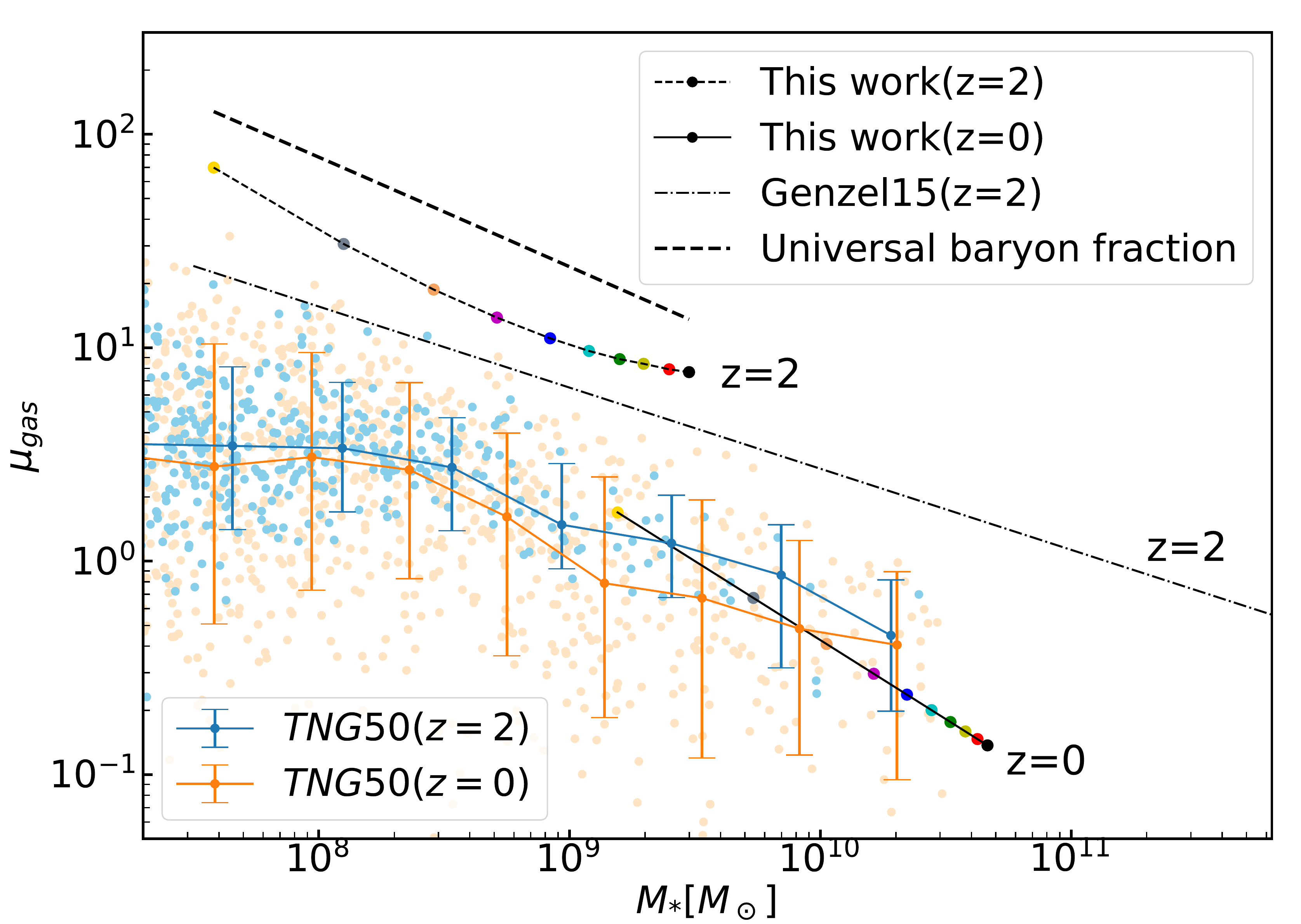}
    \caption{The gas fraction as a function of stellar mass at $z=2$ and $z=0$. The model predictions are shown as color circles connected by the black solid line ($z=0$), the black dashed line ($z=2$). The small dots are from the TNG50 simulation at $z=2$ (blue dots) and at $z=0$ (orange dots), with two lines with error-bars showing the medium and the scatter. The dash-dotted line is the best-fitting to the data at $z=2$ for the molecular gas fraction \citet{2015ApJ...800...20G}, and we extrapolate it to low-mass galaxy. The thick dashed line is a theoretical prediction by assuming that total baryon mass (star plus cold gas) in a model galaxy is the universal baryon fraction, $f_b=\frac {\Omega_b}{\Omega_m}$. It is found that the predicted gas fraction at $z=2$ is lower than the universal baryon fraction by a factor of 2, and slight higher than the data of \citep{2015ApJ...800...20G}, which should be treated as a lower limit (only molecular gas included).}
    \label{fig:fig5_fgas_observation}
\end{figure}

Using our model, we can also predict the evolution of cold gas fraction, defined as $\mu_{gas}=\frac{M_{gas}}{M_{*}}$, for the model galaxies. Using the above derived SFR for each galaxy, the cold gas amount and its evolution can be calculated from Equation \ref{eq:KS} once the SFE is specified. In principle, SFE could be a function of galaxy mass and redshift, written as $SFE(M_*,z)$. In most models, SFE is intrinsically related to the free-fall time or the dynamical time of the galaxy, see discussion in  \citet{2015ARA&A..53...51S}. 
Here we first use the fitting formula shown in Table 3 of \citet{2014A&A...564A..66B} to get the gas fraction $\mu_{gas}(z=0)$ and $M_{gas}(z=0)$. Then the value of $SFE(M_*, z=0)$ is determined using our Equation \ref{eq:KS}. It is then found that the SFE has a weak mass dependence as $SFE\sim M_{*}^{0.5}$.
This mass dependence is the same as that by \citet{2016A&A...589A..35S}.  In addition, observational work have found the redshift dependence of SFE \citep[e.g.][]{2015ApJ...800...20G,2017ApJ...837..150S,2018ApJ...860..111D,2018ApJ...853..179T}. \citet{2015ApJ...800...20G} combined the molecular gas data between $z=0-3$ and found that the gas depletion timescale $t_{depl}$ has a weak redshift dependence as $t_{depl}\sim (1+z)^{-0.3}$. As SFE is inverse to the gas depletion time, we use a combined $SFE\sim M^{0.5}_{\ast}(1+z)^{0.3}$ in our model and predict the cold gas content using Equation \ref{eq:KS}.

Figure \ref{fig:fig4_fgasevolution} shows the evolution of cold gas mass in each model galaxy. The left panel shows the amount of cold gas and the right panel is for its fraction. As one can see from the left panel that for most galaxies, the cold gas content remains almost constant during their evolution, indicating that the gas depleted by star formation and outflow is basically balanced by fresh gas infall. For massive galaxy, there is a steady decrease of cold gas with decreasing redshift, showing that the supply of cold gas is lower than the depletion. This is in general consistent theoretical prediction that quick gas accretion is maintained by cold accretion in low-mass galaxy and massive galaxy starts to quench at low redshift. The right panel shows the gas fraction as a function of redshift. It is found that almost all galaxies were gas rich at high redshift, and the fraction is higher for low-mass galaxies. For example, for the galaxy $G0$, its cold gas mass is about 50 times of its stellar mass at $z=2$.

In Figure \ref{fig:fig5_fgas_observation} we compare the predicted cold gas fraction with the TNG50 simulation \citep{2019MNRAS.490.3234N,2019ComAC...6....2N,2019MNRAS.490.3196P} and the observations at $z=0$ and $z=2$. The small scatter points are from the TNG50-1 simulation\footnote{The cold gas data here refers to the sum of molecular hydrogen and atomic hydrogen, from the calculations in \citet{2018ApJS..238...33D,2019MNRAS.487.1529D}.}, with color lines to show the median and the error-bars to show the 40 percentage from the median. The dash-dotted line is the best-fitting line by \citet{2015ApJ...800...20G} to the observed molecular gas fraction in galaxies with $M_{\ast} > 10^{9.8}M_{\odot}$ at $z=2$. We extrapolated their fit to low-mass galaxies. Our model galaxies at $z=0$ and $z=2$ are shown as filled color circles, connected by black solid line and black dotted lines, respectively. As mentioned before, our gas fraction at $z=0$ is set using the observational results by \citet{2014A&A...564A..66B}. This panel shows that the gas fraction predicted from the TNG50-1 is similar to the data at $z=0$, but there is almost no evolution with redshift in the simulation. The gas fraction at $z=2$ from our model is slightly higher than the data of \citet{2015ApJ...800...20G}. We note that the data at $z=2$ is only for molecular gas, which should be treated as a lower limit of the cold gas fraction. The thick dashed line is the universal gas fraction, which is higher than our model prediction by a factor of 2 at $z=2$. Overall, our results are consistent with the data in the sense that most galaxies were gas rich at early times.

\subsection{Constraints on the gas cycle} \label{subsec:gas_cycle}
In this section, we present our model prediction on the gas accretion rate, the recycle fraction of gas outflow by stellar feedback, and compare them with constraints from simulations. 

\subsubsection{The gas infall rate} \label{subsubsec:requried_infall}
\begin{figure}
    \centering
	\includegraphics[width=\columnwidth]{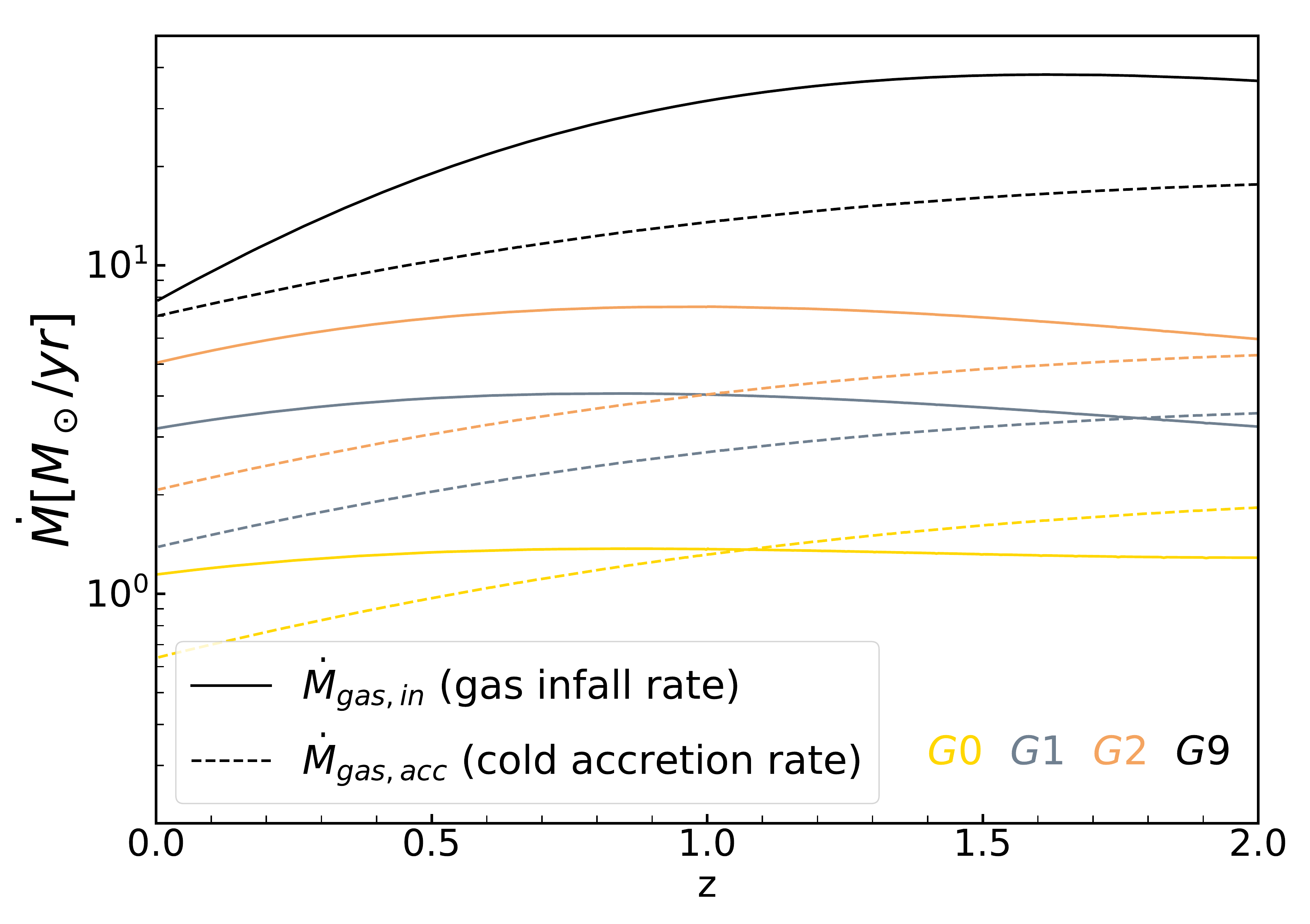}
    \caption{The evolution of the total gas infall rate (solid lines) and cold gas accretion rate due to halo growth (dashed lines) for four model galaxies. The difference between the two rates is the accretion rate from the recycle of gas outflow. It is seen that for most galaxies, the total gas infall rate is larger than the cold accretion, indicating that the gas recycle has to be effective. To make it easier to understand the label, $M_{h0}=(G0,1\times 10^{11}),\ (G1,2\times 10^{11}),\ (G2,3\times 10^{11}),\ (G9,1\times 10^{12})\ M_\odot$.}
    
    \label{fig:fig6_infallbaryon}
\end{figure}

\begin{figure*}
    \centering
	\includegraphics[scale=0.14]{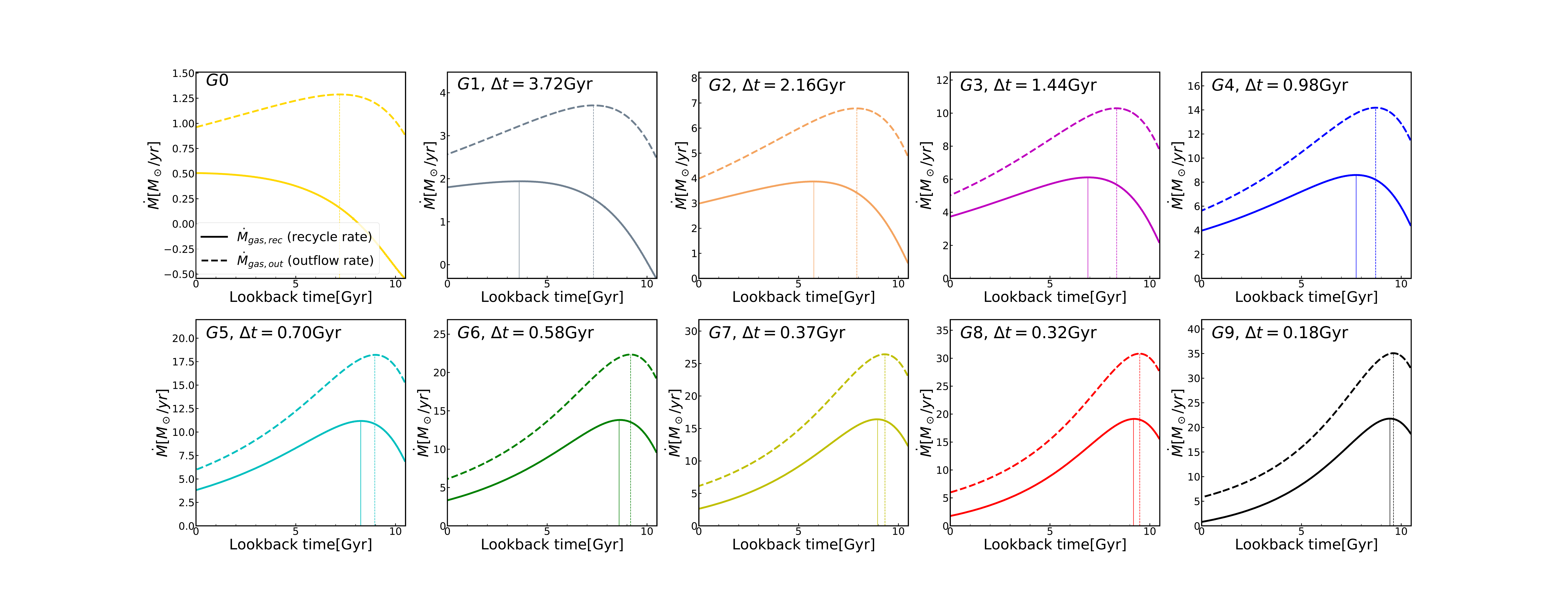}
    \caption{Evolution of gas outflow rate and recycle rate from the outflow. Each panel is for one galaxy, from G0 to G9. The dashed line is for the gas outflow rate, $\dot M_{gas,out}$, and the solid line is for gas recycle rate $\dot M_{gas,rec}$. The vertical lines mark the peak of the two lines and their time delay ($\Delta t$) can be regarded as the gas recycle time  and is labelled in each panel. To make it easier to understand the label, $M_{h0}=(G0,1\times 10^{11}),\ (G1,2\times 10^{11}),\ (G2,3\times 10^{11}),\ (G3,4\times 10^{11}),\ (G4,5\times 10^{11}),\ (G5,6\times 10^{11}),\ (G6,7\times 10^{11}),\ (G7,8\times 10^{11}),\ (G8,9\times 10^{11}),\ (G9,1\times 10^{12})\ M_\odot$.}
    \label{fig:fig7_10galaxy}
\end{figure*} 

In Section \ref{subsec:emp_method}, we mentioned that the total gas accretion rate, $\dot{M}_{gas,in}$, in a galaxy is calculated using Equation \ref{eq:total}. This accretion rate is mainly contributed by two sources. One is the gas accretion due to halo growth and the other is the recycle of gas outflow. In our selected galaxy mass range, the gas accretion due to halo growth is in the regime of cold accretion, which is fast and determined by the free-fall time. We approximate this cold accretion rate as,
\begin{equation}
    \dot{M}_{gas,acc}(t)=f_b(M_{h})\cdot \dot{M}_{h}(t-t_{ff}),
    \label{eq:baryonaccre}
\end{equation}
where $f_b(M_h)$ is the baryon fraction in the accreted dark matter halo and $t_{ff}$ is the free fall time of the galaxy. For simple consideration we set $f_b$ as the universal baryon fraction, thus $f_b(M_h) = \frac {\Omega_{b}} {\Omega_m} =0.15$, although the baryon fraction could be influenced by other effects, such as the re-ionization effect which is dependent on the mass of the accreted halo \citep[e.g.][]{2000ApJ...542..535G}. We approximate the free fall time as the dynamic time of the halo, which is calculated as  $t_{ff}=t_{dyn}=R_{vir}/V_{vir}$. The gas accretion rate from the recycle of outflow is then calculated as, 
\begin{equation}
\dot{M}_{gas,rec} = \dot{M}_{gas,in} - \dot{M}_{gas,acc}.
\label{eq:gasrecycle}
\end{equation}
In Figure \ref{fig:fig6_infallbaryon} we compare the total gas infall rate (solid lines) with the cold accretion rate (dashed lines) for four representative model galaxies, which are galaxy $G0$, $G1$, $G2$ and $G9$. It is seen that the required total gas infall rate is almost always higher than the cold accretion rate except for the low-mass galaxy, $G0$, for which the cold accretion is higher than the infall rate at $z>1.1$. For most galaxies the total infall rate is a factor of $2\sim 3$ higher than the cold accretion rate, showing that the cold accretion is not high enough to sustain the star formation in the galaxy, thus  additional supply is provided by the recycle of ejected gas. In the following section, we study the recycle process in more detail.

\subsubsection{Constraints on gas recycle from the outflow} \label{subsubsec:baryon_recycle}

\begin{figure*}
    \centering
	\includegraphics[scale=0.28]{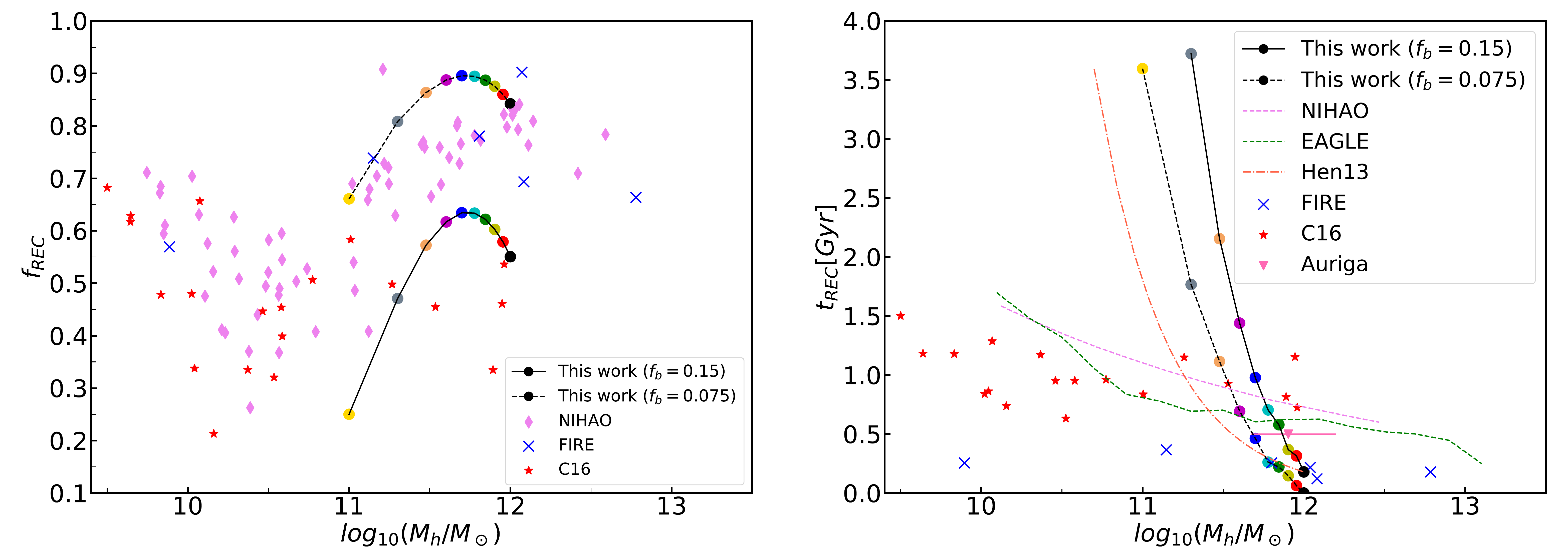}
    \caption{The total recycled mass fraction from the outflow since $z=2$ (left panel) and the recycle time (right panel). Both the black solid lines in two panels, connected with color circles, are our fiducial model predictions with $f_b=0.15$. The dashed lines with color circled connected are the model predictions if the baryon fraction in newly accreted halo is half of the universal baryon fraction, $f_b=0.075$.  All other lines or points are results from hydrodynamical simulations and the semi-analytical model, see the label and text for references. }
    \label{fig:fig8_totalfrac}
\end{figure*}

The fate of gas outflow is key for galaxy formation, but it is difficult to model. In early SAMs, it was assumed that the gas outflow is heated to the virial temperature of the halo and mixed with the hot gaseous halo for further cooling \citep[e.g.][]{2001MNRAS.328..726S}. Later version of SAMs assumed that part of the outflow remains in the halo, and part leaves the halo and is later re-incorporated in due time, both of which are dependent on halo virial velocity \citep[e.g.][]{2011MNRAS.413..101G, 2015MNRAS.451.2663H, 2015ARA&A..53...51S}. However, hydrodynamical simulations have found that the gas outflow is in multi-phase thermal state and the cycle of gas is very complicated with rich diversity among different simulations \citep[e.g.][]{2010MNRAS.406.2325O,2016ApJ...824...57C,2017MNRAS.470.4698A,2019MNRAS.485.2511T,2019MNRAS.490.4786G,2020MNRAS.497.4495M}. In this section, we compare our model constraints on the recycle of gas outflow with results from both SAMs and simulations.

We first show the evolution of the outflow rate and the gas recycle rate for our ten model galaxies in Figure \ref{fig:fig7_10galaxy}. Each panel is for one galaxy with mass increasing from the upper left to the lower right. In each panel, the vertical lines indicate the peaks for the outflow and the recycle rate. As we can see for all model galaxies, their outflow rates (dashed lines) increase with the look back time and reach a peak roughly at $8 \sim 10 Gyr$ ago. This behavior is similar to the evolution of the star formation rate shown in Figure \ref{fig:fig2_SFH} except for low-mass galaxies. For the low-mass galaxies, their star formation has not reached the peak at $z=0$, but due to their low virial velocity at high redshift, the outflow rate was also higher in the past. The solid lines are for the gas recycle rate. We note that here the recycle rate at a given time should not be considered as an instantaneous rate of gas recycled from the outflow produced at the same time, it is the recycle rate of gas from the cumulative outflow in the past.

Figure \ref{fig:fig7_10galaxy} shows two important features. One is that the recycle rate is always lower than the outflow rate, indicating that only part of the gas from the outflow is recycled to supply the star formation. We define the total recycle fraction as,
\begin{equation}
    f_{REC}=\frac {\int_{t_{z=2}}^{t_{z=0}} \dot{M}_{gas, rec}dt} {\int_{t_{z=2}}^{t_{z=0}} \dot{M}_{gas,out}dt}.
    \label{eq:f_recycle}
\end{equation}
In our model we start the evolution from the initial redshift $z=2$, and this equation gives the total recycled fraction of gas outflow by $z=0$. In fact, we find that the results are only slightly changed if we start the evolution from $z=4$. The other feature is that there is a delay between the peak of outflow rate and the recycle rate. We label the time delay in each panel, and we can see that it is small for massive galaxies and being larger for low-mass galaxies. For the lowest-mass galaxy $G0$, its peak of recycle is not reached yet at $z=0$. Although we do not model the recycle process in our model, the time delay between the two curves can be viewed as a typical recycle time of the gas outflow. This can be verified in a simple way. If we shift the two curves in each panel to overlap their peaks, we find that they have similar shapes and the ratio between their maxima is very close to the total recycle fraction calculated from Equation \ref{eq:f_recycle}. 

In Figure \ref{fig:fig8_totalfrac} we compare the predicted total gas recycle fraction and recycle time from our model with other studies. In the left panel we show the recycle fraction as a function of galaxy halo mass. The filled circles connected with a black solid line are the predictions of our fiducial model, and the other points are from a few simulation results, see the caption for references. It is seen that our results are marginally consistent with the simulation results. Both the NIHAO \citep{2019MNRAS.485.2511T} and the FIRE \citep{2017MNRAS.470.4698A} simulations indicate the recycle fraction has a peak around $M_{vir} \sim 10^{12}M_{\odot}$, sightly higher than our peak mass at around $6 \times 10^{11}M_{\odot}$. The results from \citet{2016ApJ...824...57C} do not have similar behavior, and their gas recycle fraction is between $20\%$ and $70\%$ in the plotted mass range. Overall, our gas recycle fraction is lower than the simulation results. The right panel shows the recycle time for the gas outflow. Here we include results from a few other simulations: the Auriga Simulation \citet{2019MNRAS.490.4786G}, the EAGLE simulation \citet{2020MNRAS.497.4495M}. The red dash-dotted line is the prediction by \citep{2013MNRAS.431.3373H} using the L-Galaxies model. It is found that the model, either our empirical model or the SAM of \citep{2013MNRAS.431.3373H}, over-predicts the gas recycle time for low-mass galaxies. The typical recycle times from simulations are between $0.5 Gyr$ and $1.5 Gyr$, with a weak dependence on mass than the models. Overall, this panel shows that the recycle time is short in massive galaxy. This has been pointed out by \citet{2010MNRAS.406.2325O} that the deeper gravity potential in massive galaxies make the gas recycle faster. 

In principle, hydrodynamical simulation can trace gas dynamics more accurately, thus it is very likely that both our model and the L-Galaxies model of \citet{2013MNRAS.431.3373H} do  over-estimate the gas recycle time compared to the simulation results. In our model, the high recycle time arises from the low recycle fraction of gas from the outflow. Seen from Equation \ref{eq:gasrecycle} that the gas recycle rate is the difference between the total gas infall rate and the cold accretion rate, thus our under-estimation of the gas recycle fraction is due to the over-estimation of cold accretion rate calculated from Equation \ref{eq:baryonaccre}.  Indeed, it has been well-known that the cosmic re-ionization reduces baryon fraction in low-mass halos \citep[e.g.][]{2000ApJ...542..535G}. Other than the cosmic re-ionization effect, galactic supernova feedback also disrupts the incoming cold flow and reduces the gas infall near the halo virial radius \citep[e.g.][]{2018MNRAS.474.4279M,2020MNRAS.497.4495M}. \citet{2019MNRAS.485.2511T} used the NIHAO simulation to show that galactic outflow can disrupt the infall gas up to a distance of $\sim 6 R_{vir}$ from the host galaxy, and it can reduce the baryon fraction to half of the universal one for halo with mass $\sim 3 \times 10^{10}M_{\odot}$. They named this effect as $pre-emptive\  feedback$. We do not attempt to model both the cosmic re-ionization and $pre-emptive\  feedback$ effects in this work, but  to simply illustrate how these effects will impact our results on the gas recycle process. We test it by setting the baryon fraction in all newly accreted halos as $0.5f_b$ in Equation \ref{eq:baryonaccre}. In Figure \ref{fig:fig8_totalfrac} we show this test using the black dashed lines in both panels. It is seen that decreasing the incoming baryon fraction from cold accretion will indeed result in a higher recycle fraction of gas from outflow (left panel) and a shorter recycle time (right panel). The agreement with simulations is now better, although the mass dependence in our model is still stronger than the trend in simulations.

\subsection{The mass-metallicity relation (MZR)} \label{subsec:MZR}

\begin{figure}
    \centering
	\includegraphics[width=\columnwidth]{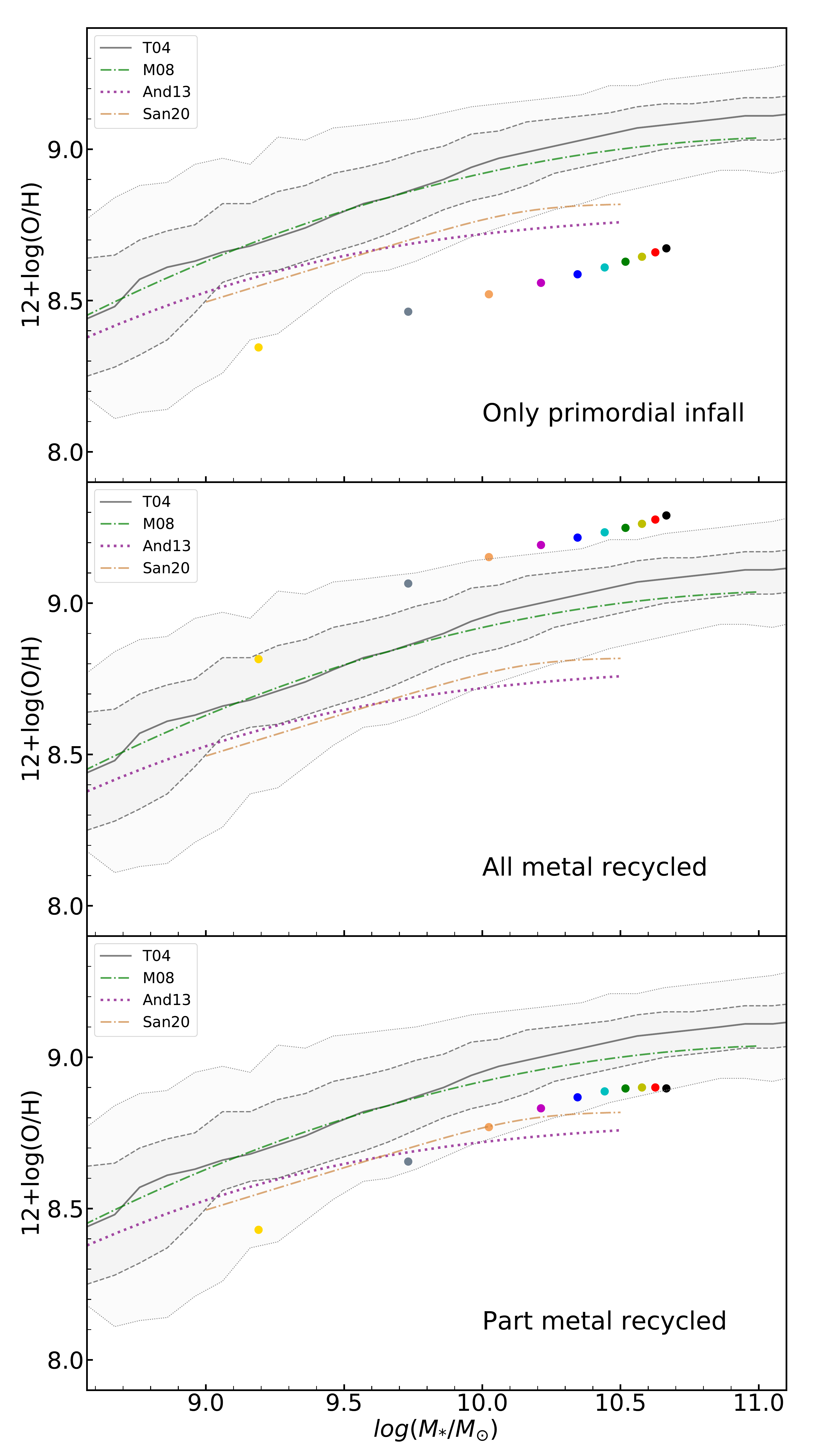}
    \caption{The mass-metallicity relation at $z=0$. The model predictions are shown as color circles. The three panels are for three different assumptions on the metallicity of the infall gas, see the text for details. All lines and shaded region are from different data:  the solid, dashed, and dotted gray lines show the median, $68\%$ contour, and $95\%$ contour, respectively, of the MZR at $z\sim 0.1$ \citep{2004ApJ...613..898T}. The green dash-dotted line is at $z=0.07$ \citep{2008A&A...488..463M}, the purple dotted line is at $z=0.027\sim 0.25$ \citep{2013ApJ...765..140A}, and the brown dash-dotted line is at $z\sim 0$ \citep{2021ApJ...914...19S}.}
    \label{fig:fig9_MZR}
\end{figure}

In this section, we compare the predicted mass-metallicity relation (MZR) with the data at $z=0$. As shown by many studies \citep[e.g.][]{2004ApJ...613..898T}, the gas-phase metallicity depends critically on the past star formation history (SFH), the gas outflow and inflow of the galaxy. Therefore, the MZR can be utilized to constrain these processes in galaxy formation and evolution. In our model, the metal content is calculated using Equation \ref{eq:metalmass1}. We note that, to proceed with that equation, one has to first set the initial metallicity of the gas. As we start the evolution of our model galaxies at $z=2$, we use the best-fitting MZR at $z=2$ by \citet{2008A&A...488..463M}, which is very similar to the recent results of \citet{2020MNRAS.491.1427S}. The metallicity of infall gas is also a free parameter, and in our model we test a few assumptions on its value, 
\begin{enumerate}
    \item The metal content in the total gas infall is set as primordial metallicity with $Z_{in} = 10^{-4}$, no matter where the infall gas is from, either the cold accretion or the recycled gas from the outflow.
    \item The metallicity of gas from cold accretion is still primordial, but all metals from the outflow are recycled instantaneously, no matter how much gas is recycled. It means that metal is actually not ejected out, but only the gas by stellar feedback.
    \item Only part of the metal in the outflow gas is recycled, and the fraction is the same as the recycled fraction of gas in the outflow.
\end{enumerate}
With the above assumptions, we can calculate the evolution of metallicity in each model galaxy. To compare with the data, we adopt the converting formula in \citet{2013MNRAS.434.1531F}:
\begin{equation}
    12+log_{10}\left(O/H\right)_{gas}=log_{10}\left(Z_{gas}/Z_\odot\right)+8.69,
    \label{eq:12OH}
\end{equation}
where $8.69$ is the solar value in terms of $12+log_{10}\left(O/H\right)$ \citep{2009ARA&A..47..481A} and $Z_\odot=0.02$.

The predicted MZR relations at $z=0$ are shown in Figure \ref{fig:fig9_MZR}, in which the three panels are for the above three different assumptions of the metal content in the infall gas, respectively. Different lines are from different observational data  \citep[e.g.][]{2004ApJ...613..898T,2008A&A...488..463M,2013ApJ...765..140A,2021ApJ...914...19S}. Our model results are shown as filled circles. The upper panel shows that the predicted MZR is below all the data if the infall gas has primordial metal content. Thus, the metal from the gas outflow must be recycled. The middle panel shows that if all metal in the outflow is recycled, the predicted MZR will be higher than the data. This indicates that metal loss must be included in the model. In the lower panel in which the recycle fraction of metal is the same as the recycled gas from the outflow, the predicted MZR is roughly consistent with the data, albeit with a slightly lower normalization. This is linked to our under-estimation of gas recycle fraction. We have tested that if a slightly higher recycle fraction about 70-80\% is used for the metal recycle, the predicted MZR agrees better with the data, but we omit the plot here. 

\section{Discussions} \label{sec:discussion}
In this section, we briefly discuss a few cases which we do not include in previous sections. One is the mass range of our model galaxy. In principle, our model can be extended to more massive halos, but as we explained before, galaxy mergers and AGN activities are becoming more effective in massive galaxies, so they hamper us to constrain the baryonic cycle from our simple empirical model. We will shortly show the predicted main sequences of massive galaxies and whether they agree with the data. Another effect to discuss is the gas outflow by stellar feedback. In our fiducial model we use the fitting formula of \citet{2015MNRAS.454.2691M} for the mass loading factor. As this is a very key parameter governing galaxy formation and most SAMs take it as a free parameter, we will discuss if our empirical model can put some constraints on this factor. 

\begin{figure}
    \centering
	\includegraphics[width=\columnwidth]{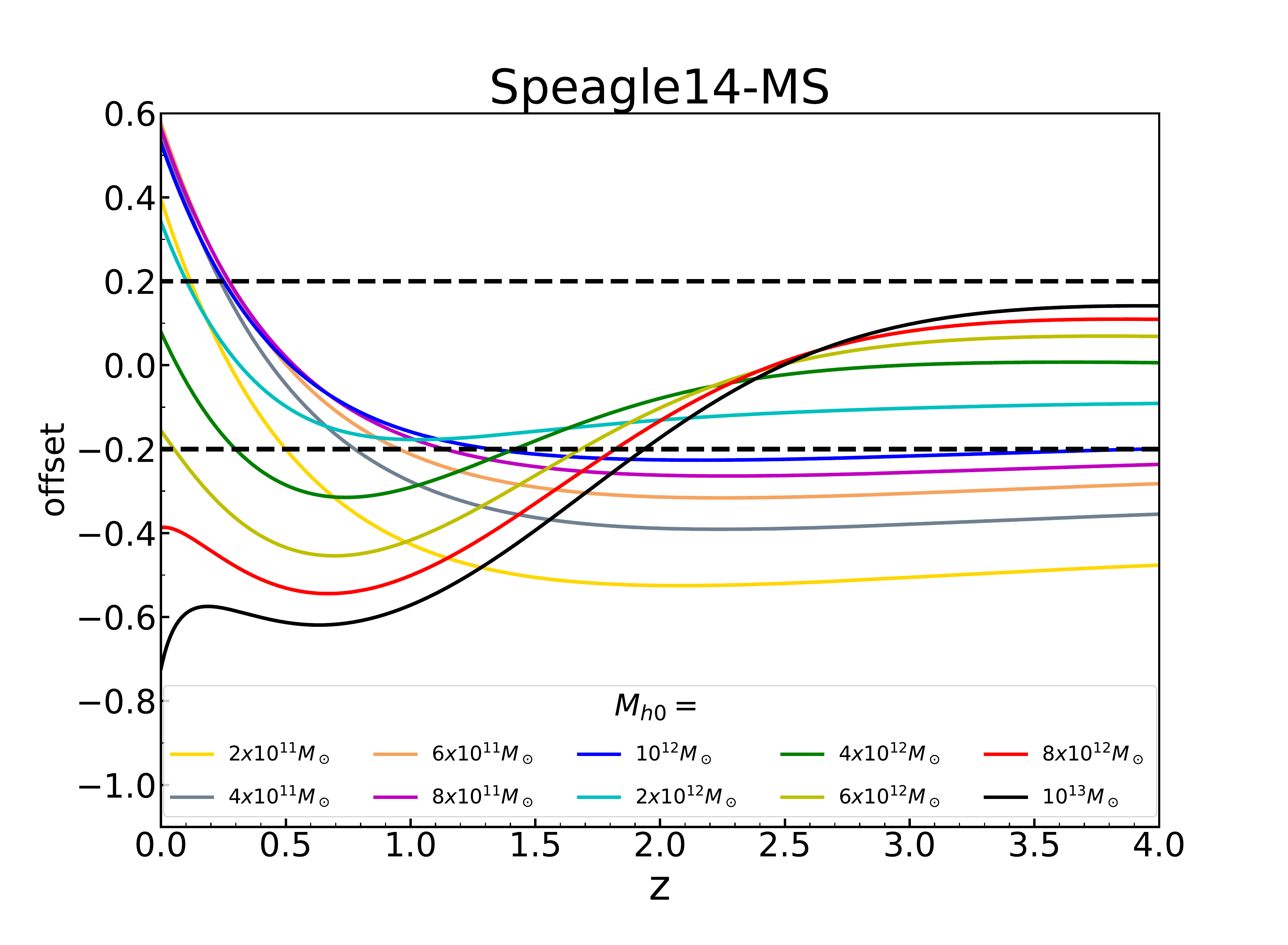}
    \caption{The offset between the predicted star formation rate and that of the observed Main Sequences by \citet{2014ApJS..214...15S} for the ten model galaxies with current ($z=0$) halo mass ranging from $2\times 10^{11}M_{\odot}$ to $10^{13}M_{\odot}$. The dashed lines indicate the typical scatter of 0.2 dex. It is clearly seen low-mass galaxies are recently becoming star forming. Massive galaxies were star forming in the past ($z>2$), and are now quenched. More massive galaxy quenched first, in agreement with the 'downsizing' scenario. Note that the upturns at $z<0.5$ for massive galaxies are artificial due to our neglect of galaxy mergers in massive haloes, see the text for details.}
    \label{fig:fig10_MS_more_sample}
\end{figure}

We have seen from Figure \ref{fig:fig3_MS} that the star formation rates of our model galaxies ($G0$-$G9$) are higher than the MS data at $z=0$, but are lower than the data at $z > 1.5$. This seems to contradict with the common perception that galaxies gradually quench at lower redshifts. In fact, evolution of a galaxy on the MS diagram depends on its mass. To demonstrate this, we select ten model galaxies with halo mass between $2\times 10^{11}M_{\odot}$ and $10^{13}M_{\odot}$, and we show the offset between the predicted star formation rate and the observed MS relations in Figure \ref{fig:fig10_MS_more_sample}. The mass of each model galaxy is labelled in the plot. Note that here we only show the comparison with the data of \citet{2014ApJS..214...15S}. The other two fitting formulae used in the paper \citep{ 2012ApJ...745..149L,https://doi.org/10.48550/arxiv.2203.10487} produce slightly different results, but the overall trends are similar. 

\begin{figure*}
    \centering
	\includegraphics[scale=0.281]{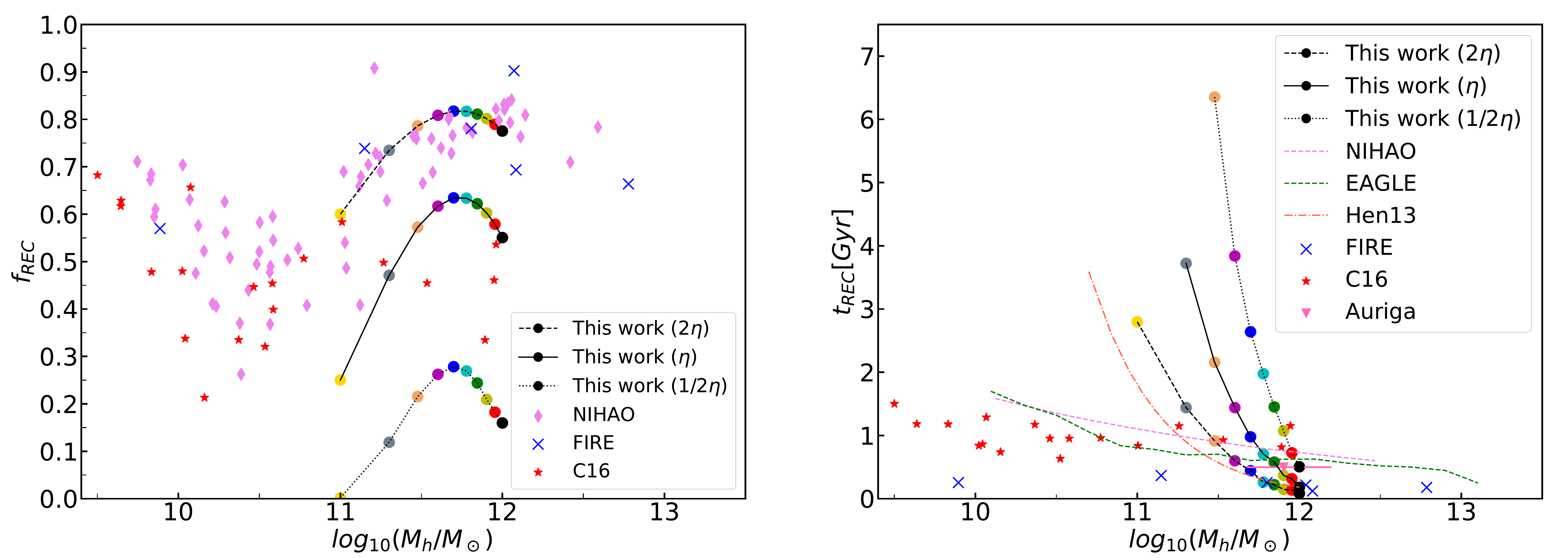}
    \caption{The same as Figure \ref{fig:fig8_totalfrac}, but here we increase/decrease the fiducial mass loading factor from \citet{2015MNRAS.454.2691M} by a factor of 2. It is found that a stronger mass loading factor, or higher stellar feedback efficiency, agrees better with the simulation results on both gas recycle fraction (dashed line with circle connected in the left panel) and recycle time (the same line style in the right panel). Our results suggest that the mass loading factor from \citet{2015MNRAS.454.2691M} could be a lower limit, and a boost by a factor of 2 is more favored by the simulation data.}
    \label{fig:fig11_change_MLF}
\end{figure*}

As we can from the plot, at $z > 2$, low mass galaxies ($M_h <10^{12}M_{\odot}$) had star formation rate lower than the MS relation and began to cross the MS at low redshift. This has already been shown in Figure \ref{fig:fig3_MS}. However, for massive galaxies ($M_h > 10^{12}M_{\odot}$) they were star forming at high redshift and more massive galaxies are more closer to the MS relation. The most massive galaxy ($M_h = 10^{13}M_{\odot}$) was actually above the MS at $z>3$. It is interesting to see that massive galaxies began to depart from the MS and became quenched at low redshift. The larger the mass is, the earlier is the quenching time. This is consistent with the 'downsizing' scenario that massive galaxies form first and quench earlier. This is also consistent with the recent studies on the evolution of the MS relation about the 'bending' mass, defined as the galaxy mass where the MS relation becomes flat, is also decreasing with redshift and it is around $M_h \sim 10^{12}M_{\odot}$ at $z=0$ \citep[e.g.][]{2013ApJ...770...57B, 2022arXiv220310880D}. This 'bending' mass is interpreted as the transitional mass where gas accretion is switched from cold accretion to hot accretion. We note that massive galaxies in our model show upturns of their star formation rates relative to the MS relations at $z<1$. This behavior is artificial. We have mentioned that we neglected galaxy mergers in our model, so all stellar mass growth is ascribed to star formation which might be over-estimated. \cite{2009ApJ...696..620C} and \cite{2013MNRAS.428.3121M} also find that for massive galaxies at low redshifts, the contribution of galaxy mergers to the stellar mass growth of galaxies is non-negligible.

Now we discuss the effect of changing the mass loading factor, $\eta$. The mass loading factor we used in Equation \ref{eq:massloadingfactor} is from the best-fitting relation by \citet{2015MNRAS.454.2691M} to the FIRE simulations. \citet{2020MNRAS.494.3971M} have compared the mass loading factor from EAGLE simulation with other hydrodynamical simulations and different versions of SAM. They found large discrepancy between simulations and SAMs, mainly due to different definitions of outflow and physics implemented in different simulations. Overall, the mass loading factor from EAGLE \citep{2020MNRAS.494.3971M} is larger than the Horizon-AGN simulation \citep{2017MNRAS.472..949B}, but lower than FIRE simulation \citep{2014MNRAS.445..581H}, TNG50 simulation \citep{2019MNRAS.490.3234N} and the NIHAO simulation \citep{2019MNRAS.485.2511T}. Among these hydrodynamical simulations, the mass loading factor in NIHAO is the strongest. We refer the readers to the paper by \citet{2020MNRAS.494.3971M} for more details. In Figure \ref{fig:fig11_change_MLF} we show two additional cases where we increase/decrease the mass loading factor, $\eta$, by a factor of 2. It is interesting to find that increasing the stellar feedback efficiency, or the mass loading factor, results in a higher recycle fraction from the outflow (dashed line in the left panel) and a shorter recycle time (dashed line with circles in the right panel). Reducing the stellar feedback brings an opposite effect (dotted lines with filled circles). Using our empirical model introduced in Section \ref{subsec:emp_method}, it is straightforward to show that the recycle fraction is related to the outflow as, 
\begin{equation}
    \begin{aligned}
    f_{REC} &= \frac{\dot M_{gas,rec}}{\dot M_{gas, out}}\\
    &= \frac{\dot M_{gas,in}-\dot M_{gas,acc}}{\dot M_{gas, out}}\\
    &=\frac{(\dot M_{gas}+\dot M_{*}+\dot M_{gas,out})-\dot M_{gas,acc}}{\dot M_{gas,out}}\\
    &=1-\frac{\dot M_{gas,acc}-\dot M_{gas}-\dot M_*} {\eta \dot M_{*}/(1-R)}.
    \end{aligned}
	\label{eq:MLF-RE}
\end{equation}
In our model, the cold accretion rate (Equation \ref{eq:baryonaccre}), cold gas and stellar mass are all independent of the parameter $\eta$, thus it is understandable from Equation \ref{eq:MLF-RE} that a larger $\eta$ will result in a larger $f_{REC}$ and a lower recycle time. 

In general, most hydrodynamical simulations have implemented different stellar feedback physics and tuned their parameter to best fit different set of observational data, it is naturally to see there is a large diversity among their predictions on gas recycle. As we mentioned before, although the mass loading factor is largely diverse in different simulations  \citep{2020MNRAS.494.3971M}, the predicted gas recycle fraction and the gas recycle time from the outflow are not significantly different (points in Figure \ref{fig:fig11_change_MLF}). One possible reason is that gas recycle fraction does not depend strongly on the details of the stellar feedback, but the dynamical properties of the halo where a galaxy stays. Our results in Figure \ref{fig:fig11_change_MLF} indicate that the mass loading factor we used \citep{2015MNRAS.454.2691M} could be a lower limit. In fact, our results favor a stronger mass loading factor and a recycle time which are similar to the constraints from the L-Galaxies model \citep{2013MNRAS.431.3373H}. 

\section{Summary} \label{sec:summary}
In this paper, we use the evolution of the galaxy stellar-to-halo mass relation to obtain star formation history of a few model galaxies with halo mass ranging from $10^{11}M_{\odot}$ to $10^{13}M_{\odot}$ at $z=0$. We combine their star formation history with a simple empirical model for galaxy evolution to constrain the gas cycle. Our results can be summarized as the followings, 

\begin{enumerate}
    \item Galaxies are not always evolving along the main sequences (MS) throughout their life times. Low-mass galaxies ($M_h <10^{12}M_{\odot}$ at $z=0$) had star formation rate lower than the MS at $z>1$, and they gradually crossed the MS at $z<0.5$ and stay above the MS at $z=0$. High-mass galaxies ($M_h >10^{12}M_{\odot}$ at $z=0$) were star forming galaxies at $z>2$ with star formation rate close to the MS relations. With decreasing redshift, massive galaxies began to depart from the MS and gradually quenched. More massive galaxies quenched earlier. The evolution of galaxy star formation history generally agrees with the 'downsizing' scenario that massive galaxy forms first and is quenched early.
    
    \item By incorporating a star formation efficiency with dependence on both stellar mass and redshift as $SFE \sim M_{\ast}^{0.5}(1+z)^{0.3}$, we find that the cold gas fraction in all model galaxies at $z=2$ is lower than the universal baryon fraction by a factor of 2, and it is consistent with the observations. The predicted cold gas fraction from TNG50-1 simulation does not evolve with redshift and is lower than the data at $z=2$.
    
    \item It is found that, in addition to the cold gas accretion due to halo growth, gas recycle from outflow is required to sustain the star formation in our model galaxy. The recycle fraction is between 20\% and 60\% for galaxy with current halo mass between  $10^{11}M_{\odot}$ and $10^{12}M_{\odot}$. The peak recycle fraction is at halo mass around $6\times 10^{11}M_{\odot}$ and low-mass galaxy has lower recycle fraction. This is quantitatively consistent with simulation results. The gas recycle time for massive galaxy is around $0.5 Gyr$, in consistent with simulation results, but it is about $4 Gyr$ for low-mass galaxy, higher than the simulation results. 
    
    \item We predicted the mass-metallicity relation using our model. In the model, the metal in inter-stellar medium (ISM) is enriched by newly formed stars. The metal could be replenished by newly accreted gas or diminished by gas outflow. It is found that if all metal in the outflow is lost and infall gas has primordial gas metallicity, $Z_{gas}\sim 10^{-4}$, the predicted metallicity is lower than the data at $z=0$ even we have tuned the initial mass-metallicity relation to match the data at $z=2$. However, if all metal in the outflow is recycled to the ISM, the predicted MZR is higher than the data at $z=0$ and can be excluded. An equal fraction of recycle for both gas outflow and metal produces a MZR that agrees with the data slightly better.

\end{enumerate}

We also find that both the gas recycle fraction and recycle time are affected by the processes of cold gas accretion and stellar feedback. From the discussion of Section \ref{subsubsec:baryon_recycle} and Figure \ref{fig:fig8_totalfrac}, we can conclude that if the gas fraction in newly accreted halo is lower than the universal baryon fraction (e.g. $f_b=0.075$), as indicated by previous simulation results \cite[e.g.][]{2018MNRAS.474.4279M, 2019MNRAS.485.2511T}, the required gas recycle fraction from the outflow is higher and the recycle time is shorter, which are more consistent with the simulation results shown in Figure \ref{fig:fig8_totalfrac}. From the discussion on the effect of stellar feedback in Section \ref{sec:discussion} and Figure \ref{fig:fig11_change_MLF}, we find that a stronger feedback efficiency or a higher mass loading factor than that from the FIRE simulation \citep{2015MNRAS.454.2691M} will lead to the same result as above. So in our model, the cold accretion rate and mass loading factor can be constrained by the cooling of the recycled gas .

It has been long recognized that the circular galactic medium (CGM) of a galaxy is composed of gas from both incoming accretion and ejected gas from star formation, and CGM plays an important role in governing galaxy formation and evolution \citep[e.g.][]{2017ARA&A..55..389T}.Great efforts have been paid to observe the CGM and characterize its fundamental properties (e.g total mass, cooling rate). For example, \cite{2021ApJ...916..101Z} recently measured the mass cooling rates for CGM gas as a function of galactic stellar mass using the observations of $H\alpha$ emission line. Simply considering the galactic gas outflow forms the major part of the CGM of a galaxy \citep{2017ARA&A..55..389T}, the mass cooling rate of CGM gas can be greatly simplified as the mass recycle rate of outflow gas ($\dot{M}_{gas,rec}$). Then using our Equation \ref{eq:MLF-RE}, with the recycle fraction ($f_{REC}$) from simulations, the mass loading factor $\eta$ will be uniquely determined. Therefore, the cold accretion rate for galaxies ($\dot{M}_{gas,acc}$) will also be uniquely determined. In addition, with the mass cooling rate of CGM gas and the total mass of CGM gas, the cooling time can be roughly estimated. Of course, the above intuitive derivation is based on a few simple assumptions, and more accurate and complex models are needed to realize and test it. Nevertheless, by this simple derivation we can see the important role that future observation of CGM observations can greatly constrain the baryon cycle process in galaxy formation.

On the other hand, it would also be interesting to incorporate the constraints on gas recycle fraction and recycle time obtained in this study into a semi-analytical model to see if it can reproduce the global  properties of galaxy, such as the evolution of galaxy stellar mass function and star formation main sequences. We will investigate them in a future work.

\section*{Acknowledgements}
This work is partly supported by the National Key Research and Development Program (No. 2022FYA1602903), NSFC (No.11825303, 11861131006), the Fundamental Research Fund for Chinese Central Universities (No.226-2022-00216), the science research grants
from the China Manned Space project with NO.CMS-CSST-2021-A03, CMS-CSST-2021-B01 and the cosmology simulation database (CSD) in the National Basic
Science Data Center (NBSDC-DB-10). We thank Miao Li, Yu Luo, Qingxin Wen, Yisheng Qiu and Liping Wei for helpful discussions.

\section*{Data Availability}
The data underlying this article will be shared on reasonable request to the corresponding author.



\bibliographystyle{mnras}
\bibliography{reference} 

\bsp	
\label{lastpage}
\end{document}